\documentstyle[aps,pra]{revtex}
\begin{document}
\newcommand{\pp}[1]{\phantom{#1}}
\newcommand{\be}{\begin{eqnarray}}
\newcommand{\ee}{\end{eqnarray}}
\newcommand{\ve}{\varepsilon}
\newcommand{\vs}{\varsigma}
\newcommand{\vp}{\varphi}
\newcommand{\Tr}{{\rm Tr\,}}
\newtheorem{th}{Theorem}
\newtheorem{lem}[th]{Lemma}

\title{
Manifestly covariant formulation of discrete-spin and real-mass\\ 
unitary representations of the Poincar\'e group
}
\author{Marek Czachor}
\address{
Katedra  Fizyki Teoretycznej i Metod Matematycznych 
\\
 Politechnika Gda\'{n}ska,
ul. Narutowicza 11/12, 80-952 Gda\'{n}sk, Poland
}
\maketitle
\begin{abstract}
Manifestly covariant
formulation of discrete-spin, real-mass unitary representations of
the Poincar\'e group is given.
We begin with
a field of spin-frames 
associated with 4-mometa $p$ and use them to
simplify the eigenvalue problem for the Pauli-Lubanski 
vector projection in
a direction given by a world-vector $t$. As opposed to the
standard treatments where $t$ is a constant time direction,
our $t$ is in general $p$-dependent and timelike,
spacelike or null. The
corresponding eigenstates play a role of a basis used to
define Bargmann-Wigner spinors which form a carrier space of the
unitary representation.
The construction does not use the Wigner-Mackey induction procedure,
is manifestly covariant and works simultaneously in
both massive and massless cases (in on- and off-shell versions).
Of particular interest are special Bargmann-Wigner spinors ($\omega$-spinors)
associated with flag pole directions of the
spin-frame field $\omega_A(p)$.
\end{abstract}

\section{Introduction}
Unitary representations of the Poincar\'e group\footnote{By 
the Poincar\'e group I mean the semidirect
product of 4-translations in Minkowski space and $SL(2,C)$.} 
are typically
given in a form which is not manifestly covariant. 
One often speaks of {\it noncovariant\/} or {\it Wigner wave
functions\/} , 
which belong to a carrier space of a unitary representation,
and {\it covariant\/} or {\it spinor wave functions\/} which
belong to a nonunitary representation~\cite{Wig,BR,Ohnuki,kim1}. 
The covariant wave functions do not have a natural probability
interpretation although there exists a nonunitary transformation
between them and Wigner states. If one incorporates this
transformation into a scalar product one obtains a form which by
some authors \cite{N1,N2,N3} is called manifestly covariant.
From our perspective this ``manifest" covariance is not manifest
as it involves a dependence on fixed 4-momenta
used for induction of the representation and does not allow for an explicit
(abstract) index formulation. A manifestly covariant formulation
of unitary representations of the Lorentz group has been
recently discussed in \cite{MS}. 

The unitary representations discussed in literature are usually
irreducible. In the context of the Poincar\'e group this means
they are characterized by definite values of spin and mass. 
Although essentialy there is no physical problem with 
concrete values of spin, a concrete value of mass leads to
practical and formal difficulties. The most obvious
example is the self-energy mass
correction which shows that mass of an interacting field is a
dynamical object which has to be renormalized. 
Also at a purely formal level there are reasons to replace the 
concrete-mass (or ``on-shell") formalism with the off-shell one 
\cite{H1,H2,H3,H4,H5,H6,arg1,arg2,arg3,Gos}. 
In the context of this work the problem we face is the question
of {\it manifest\/} covariance in momentum representation: 
To have a manifestly covariant formulation we have to use
four components of $p$. But since the mass hyperboloid is a
three dimensional manifold, we express the zeroth
component of $p$ as $p_0=\pm\sqrt{\bbox p^2 + m^2}$ and in this
way introduce a preferred three-momentum reference
frame.  This preferred frame becomes manifest whenever we
explicitly write generators of a representation in
question~\cite{Ohnuki,gen1,gen2}. Obviously, this is typical of {\it
any\/} representation (unitary or not) and the so-called
covariant wave functions are not, in this sense, manifestly
covariant either. This formal difficulty is related to the old
problem of relativistic localization and relativistic position
operator~\cite{BR,pos1,pos2,pos3,pos4,pos5,IBB1,IBB2}.

The difficulties with the latter problem motivated 
myself to look for a manifestly covariant
reformulation of unitary representations of the Poincar\'e group. 
It turns out that there are several different levels where the
noncovariance is rooted. The one with the on-shell noncovariance
cannot be overcome unless we switch to an off-shell formulation.
Therefore  I generally write formulas in a form
which enables us to use them in both on- and off-shell versions.
The delicate point with the off-shell formulation is the
massless boundary $m=0$ which is taken care of in detail.

Other levels of noncovariance can be removed by the method of
null frames I use. This allows us to treat the
massive and massless cases 
on the same footing 
and obtain formulas which either depend
on $p\cdot p=m^2$ in an explicit and nonsingular way, or are
mass independent. The latter is made possibile by a nonstandard
choice of spin eigenstates, namely those corresponding to the
Pauli-Lubanski vector projections in {\it null\/} and
$p$-dependent directions defined by
flag poles of a specifically chosen field of spin-frames. 
In this way we circumvent the implicit
noncovariance introduced by inducing from little groups of
preferred four-momenta. The only price we pay for the
manifest covariance is the $SL(2,C)$-spin-reducibility of
representations on the massless boundary. 

The choice of $p$-dependent directions frees us of the
noncovariance associated with the $\bbox p$-dependent 
helicity amplitudes. From our perspective 
the helicity eigenstates are noncovariant in two ways. They
are either projections of the three-spin on the three-momentum $\bbox p$,
or correspond to the zeroth-component of the Pauli-Lubanski four-vector. 

The null directions and the spin-frames associated with them appear
in my formalism quite naturally as a means of unifying the
massive and massless cases into a single framework. 
The corresponding amplitudes seem to have been overlooked
in the representation theory of the Poincar\'e group although
they appeared implicitly in the context of geometric
quantization of spinor fields. So as an interesting by-product
of the manifestly covariant formulation we arrive at a
physical and group representation 
interpretation of some purely formal objects used in
geometric quantization. One should mention here that null
hyperplanes are occasionally used to define position space
scalar products for the Dirac equation \cite{Gitman} but there
is no direct relationship of such a null formalism to the one
described below. 

The layout of the paper is as follows. In Sec.~\ref{Sec.II} I present
a spinor formulation of the eigenvalue problem for the
Pauli-Lubanski vector. I begin with a separate treatment of the
massless and massive cases, introduce the
four-momentum-dependent spin-frames, 
and finally use them to derive a simple form which looks
the same for both $m=0$ and $m\neq 0$. I also introduce the basic bispinor
projectors. Their essential algebraic and transformation
properties are discussed in Sec.~\ref{Sec.III}. The role of the
projectors is explained in Sec.~\ref{Sec.IV} where their
relationship to the off- and on-shell Bargmann-Wigner
equations is made clear. The representations are unitary with
respect to the off- and on-shell scalar products discussed in
Sec.~\ref{Sec.V}.  In Sec.~\ref{Sec.VI} I introduce 
normalized eigenvectors and associated amplitudes. 
The ``null" amplitudes, called $\omega$-spinors,
lead to a very simple and elegant formalism.
In Sec.~\ref{Sec.VII} I discuss in detail the 
transformation properties of $\omega$-spinors and use them 
to formulate the unitary representations in a
manifestly covariant way. In Sec.~\ref{Sec.VIII} a generalized
off-shell position representation on the Poincar\'e group is briefly
discussed. 

The bispinor convention I use is explained in
Appendix~\ref{bispinors}. It is analogous to the standard twistor
notation but, as opposed to twistors which are always ``unprimed", I
found it very useful to introduce primed bispinor 
indices. The only exeption with respect to the standard
tensor-spinor notation is in transvection of two world-vectors,
which I usually denote by a dot, i.e. $x\cdot p=x^ap_a$, etc.

\section{
Eigenvalue problem for the Pauli-Lubanski vector in spinor form}
\label{Sec.II}

Spin eigenstates that span carrier spaces of unitary represenations
of the Poincar\'e group discussed in literature correspond to
projections of the P-L vector in either timelike or spacelike
directions. It turns out that the formalism becomes the simplest
if one takes {\it null\/} directions. Their
choice naturally follows from a spinor formulation of the
P-L vector eigenvalue problem. 

Consider the Poincar\'e Lie algebra whose elements are $P^a$ and
${J}^{ab}$. 
The Pauli-Lubanski (P-L) 
vectors corresponding to $(1/2,0)$ and $(0,1/2)$ spinor
representations of $SL(2,C)$ are 
\be
S^a{_{X}}{^{Y}}&=&P_b{^*}{J}^{ab\pp A Y}_{\pp {aa}X},\\
S^a{_{X'}}{^{Y'}}&=&P_b{^*}{J}^{ab\pp A Y'}_{\pp
{aa}X'},
\ee
where the asterisk denotes dualization \cite{PR}.
Their momentum representation is 
\begin{eqnarray}
S^a(p){_{X}}{^{Y}}&=&
-\frac{1}{2}\Bigl(p{_{X}}{^{A'}}\varepsilon^{AY}
-\varepsilon{_{X}}{^A}p^{YA'}\Bigr),
\\
S^a(p){_{X'}}{^{Y'}}&=&
\frac{1}{2}\Bigl(p{^{A}}{_{X'}}\varepsilon^{A'Y'}
-\varepsilon{_{X'}}{^{A'}}p^{AY'}\Bigr).
\end{eqnarray}

\subsection{$m=0$ representations}

\subsubsection{P-L vector in terms of spin-frames}

Let $p^a=\pi^A\bar
\pi^{A'}$ and $\omega^A$ be a spin-frame partner of 
$\pi^A$, i.e. $\omega_A\pi^A=1$. 
Then 
\begin{eqnarray}
S^a(p){_{X}}{^{Y}}&=&
-\frac{1}{2}\Bigl(\pi{_{X}}\varepsilon^{AY}
-\varepsilon{_{X}}{^A}\pi^{Y}\Bigr)\bar \pi{^{A'}}
\\
S^a(p){_{X'}}{^{Y'}}&=&
\frac{1}{2}\Bigl(\bar \pi{_{X'}}\varepsilon^{A'Y'}
-\varepsilon{_{X'}}{^{A'}}\bar \pi^{Y'}\Bigr)\pi{^{A}}
\end{eqnarray}
and 
\begin{eqnarray}
S^a(p){_{X}}{^{Y}}\pi{_{Y}}&=&
-\frac{1}{2}p^a\pi{_{X}}\label{pi1}
\\
S^a(p){_{X'}}{^{Y'}}\bar \pi{_{Y'}}&=&
\frac{1}{2}p^a\bar \pi{_{X'}}\label{pi2}\\
S^a(p){_{X}}{^{Y}}\omega{_{Y}}
&=&
+\frac{1}{2}p^a\omega{_{X}}
-
\pi_X\omega^A
\bar \pi{^{A'}}
\\
S^a(p){_{X'}}{^{Y'}}\bar \omega{_{Y'}}&=&
-\frac{1}{2}p^a\bar \omega{_{X'}}
+
\bar \pi_{X'}\bar \omega^{A'}
\pi{^{A}}
\end{eqnarray}
Equations (\ref{pi1}), (\ref{pi2}) solve the eigenvalue problem.
For higher spin $(M/2,N/2)$ fields we find 
\be
S^a(p){_{\cal X}}{^{\cal Y}}\pi{_{\cal Y}}&=&
-\frac{1}{2}\Bigl(M-N\Bigr)
\pi{_{\cal X}}
p^a
\ee
where ${\cal X}=X_1\dots X_MX'_1\dots X'_N$, 
$\pi_{\cal X}=\pi_{X_1}\dots \pi_{X_M}\bar \pi_{X'_1}\dots \bar
\pi_{X'_N}$, etc.

\subsubsection{Projections of P-L vector in $\omega$-direction}

The projections $S(\omega,p)=\omega^aS_a$ 
of the P-L vector in the  $\omega^A\bar \omega^{A'}$ direction are
\be
S(\omega,p){_{A}}{^{B}}&=&
\frac{1}{2}\Bigl(
\pi_{A}\omega^{B}+ \omega_{A}\pi^{B}\Bigr),\\
S(\omega,p){_{A'}}{^{B'}}&=&-
\frac{1}{2}\Bigl(
\bar \pi_{A'}\bar \omega^{B'}+ 
\bar \omega_{A'}\bar \pi^{B'}\Bigr)
\ee
implying
\be
S(\omega,p){_{A}}{^{B}}\omega_{B}&=&
\frac{1}{2}\omega_{A},\label{pl000}\\
S(\omega,p){_{A'}}{^{B'}}\bar \omega_{B'}&=&
-\frac{1}{2}\bar \omega_{A'}.\label{pl000'}\\
S(\omega,p){_{A}}{^{B}}\pi_{B}&=&
-\frac{1}{2}\pi_{A},\label{pl00}\\
S(\omega,p){_{A'}}{^{B'}}\bar \pi_{B'}&=&
\frac{1}{2}\bar \pi_{A'},\label{pl00'}
\ee
\subsection{$m\neq 0$ representations}

The massive case is more complicated since the components of the
P-L vector no longer commute. 

\subsubsection{Projections of P-L vector in a general $t$-direction}

Consider a projection
of the P-L
 vector in the direction of a (timelike, spacelike or
null, and generally $p$-dependent) world-vector $t$ 
\be
S(t,p){_{X}}{^{Y}}=
\frac{1}{2}\Bigl(t{^{Y}}{_{X'}}p{_{X}}{^{X'}}
+t{_{XX'}}p^{YX'}\Bigr),\\
S(t,p){_{X'}}{^{Y'}}=
-\frac{1}{2}\Bigl(t{_{X}}{^{Y'}}p{^{X}}{_{X'}}
+t{_{XX'}}p^{XY'}\Bigr).
\ee
Notice that 
\be
\overline{S(t,p){_{X}}{^{Y}}}=-S(t,p){_{X'}}{^{Y'}}\label{cc-of-S}
\ee
so that complex conjugation reverses the sign of spin. 
Eigenvalues of a symmetric spinor $S_{XY}$  are $\pm \bigl[
-\frac{1}{2}S_{AB}S^{AB}\bigr]^{1/2}$. 
An analogous formula holds for symmetric spinors with two primed
indices. 

Therefore the projection of the
P-L vector in the direction $t$  has eigenvalues
\be
\frac{1}{2}\lambda^{(\pm)}(t,p)
&=&\pm \frac{1}{2}\sqrt{(t\cdot p)^2 -m^2t^2 },\label{plev}
\ee
where $m^2=p\cdot p$ and $t^2=t\cdot t$. 
Formula (\ref{plev}) shows that there exists a privileged choice
of $t$, namely {\it null\/} and pointing in the direction of
$p$ (in the sense of pointing into the future or the past) since
in this case 
\be
 \frac{1}{2}\lambda^{(\pm)}(t,p)
&=&\pm \frac{1}{2}t\cdot p \label{plev'}
\ee
which is analogous to the massless case even though, in general,
$m^2\neq 0$ in (\ref{plev}). 

\subsubsection{Projectors associated with the eigenvectors}

The corresponding eigenvectors are
determined by the projectors
\be
\Pi^{(\pm)}(t,p){_A}{^B}&=&
\frac{1}{2}\Bigl(
\varepsilon{_A}{^B} + \frac{2}{\lambda^{(\pm)}}
S(t,p){_A}{^B}\Bigr),\\
\Pi^{(\pm)}(t,p){_{A'}}{^{B'}}&=&
\frac{1}{2}\Bigl(
\varepsilon{_{A'}}{^{B'}} + \frac{2}{\lambda^{(\pm)}}
S(t,p){_{A'}}{^{B'}}\Bigr).
\ee
The ``sign-of-energy" projectors are defined as 
(see Appendix~\ref{bispinors})
\be
{\Pi}_\pm (p){_\alpha}{^\beta}=
{\Pi}_\mp (-p){_\alpha}{^\beta}=
{\Pi}(\pm p){_\alpha}{^\beta}=
\frac{1}{2}
\left(
\begin{array}{c}
\varepsilon{_A}{^B} \\
\pm\sqrt{\frac{2}{p\cdot p}} p{_{A}}{^{B'}}\\
\mp\sqrt{\frac{2}{p\cdot p}} p{^{B}}{_{A'}} \\
\varepsilon{_{A'}}{^{B'}}
\end{array}
\right).\label{note}
\ee
They commute with the
P-L vector corresponding to the bispinor $(1/2,0)\oplus(0,1/2)$
representation: 
\be
{\Pi}_\pm (p){_\alpha}{^\beta}{S}^a(p){_\beta}{^\gamma}=
{S}^a(p){_\alpha}{^\beta}{\Pi}_\pm (p){_\beta}{^\gamma},
\ee
where
\be
{S}^a(p){_\alpha}{^\beta}=
\left(
\begin{array}{c}
S^a(p){_A}{^B} \\
 0\\
0 \\
 S^a(p){_{A'}}{^{B'}}
\end{array}
\right).
\ee
Let
\be
{\Pi}^{(\pm)}(t,p){_{\alpha}}{^{\beta}}=
\left(
\begin{array}{c}
{\Pi}^{(\pm)}(t,p){_{A}}{^{B}} \\
 0\\
0 \\
 {\Pi}^{(\pm)}(t,p){_{A'}}{^{B'}}
\end{array}
\right).
\ee
Then
\be
{\Pi}^{(\pm)}(t,p){_{\alpha}}{^{\beta}}
{\Pi}_\pm (p){_\beta}{^\gamma}&=&
{\Pi}_\pm (p){_\alpha}{^\beta}
{\Pi}^{(\pm)}(t,p){_{\beta}}{^{\gamma}}=
{\Pi}^{(\pm)}_{\pm}(t,p){_{\alpha}}{^{\gamma}}\label{PPi}
\nonumber\\
&=&
\frac{1}{4\lambda^{(\pm)}}
\left(
\begin{array}{c}
\lambda^{(\pm)}\varepsilon{_A}{^C}
+ t{^C}{_{X'}}p{_{A}}{^{X'}} + t{_{AX'}}p{^{CX'}}\\
\pm\sqrt{\frac{2}{p\cdot p}}\bigl[(\lambda^{(\pm)}-t\cdot p) 
p{_{A}}{^{C'}}+ m^2 t{_{A}}{^{C'}}\bigr]\\
\mp\sqrt{\frac{2}{p\cdot p}}\bigl[(\lambda^{(\pm)}+t\cdot p) 
p{^{C}}{_{A'}}- m^2 t{^{C}}{_{A'}}\bigr]\\
\lambda^{(\pm)}\varepsilon{_{A'}}{^{C'}}
- t{_X}{^{C'}}p{^{X}}{_{A'}} - t{_{XA'}}p{^{XC'}}
\end{array}
\right)
\ee
where $\lambda^{(\pm)}=\lambda^{(\pm)}(t,p)$. The signs ``$\pm$"
of energy are independent of the signs ``$(\pm)$" of spin. 

\subsubsection{The projectors in terms of spin-frames ---
analogy with $m=0$}

To simplify the form of (\ref{PPi}) 
consider the decomposition 
\be
p^a=\pi^{a}+\frac{m^2}{2}\omega^{a}
=
\pi^{A} \bar \pi^{A'}
+
\frac{m^2}{2}
\omega^{A} \bar \omega^{A'}
,\label{p}
\ee
$\omega_{A}\pi^A=1$, 
of a (timelike or null) future-pointing world-vector
$p^a$ (see Appendix~\ref{spin-frame}). 

Let us take $t=\omega$. 
The eigenvalues of $S(\omega,p)$ are 
$\pm 1/2$
and the corresponding projectors are
\be
{\Pi}^{(+)}_{\pm}(\omega,p){_{\alpha}}{^{\gamma}}
&=&
\frac{1}{2}
\left(
\begin{array}{c}
\omega{_{A}}\pi{^{C}}\\
\pm\sqrt{\frac{p\cdot p}{2}}
 \omega{_{A}}\bar \omega{^{C'}}\\
\mp\sqrt{\frac{2}{p\cdot p}}
\bar \pi{_{A'}}\pi{^{C}}\\
- \bar \pi{_{A'}} \bar \omega{^{C'}}
\end{array}
\right)\label{30}
\\
{\Pi}^{(-)}_{\pm}(\omega,p){_{\alpha}}{^{\gamma}}
&=&
\frac{1}{2}
\left(
\begin{array}{c}
- \pi{_{A}}\omega{^C}\\
\pm\sqrt{\frac{2}{p\cdot p}}\pi{_{A}}\bar \pi{^{C'}}\\
\mp\sqrt{\frac{p\cdot p}{2}}
\bar \omega{_{A'}}\omega{^{C}}\\
\bar \omega{_{A'}}\bar \pi{^{C'}}
\end{array}
\right).\label{31}
\ee
We obtain also formulas analogous to the massless case:
\be
S(\omega,p){_{A}}{^{B}}&=&
\frac{1}{2}\Bigl(
\omega_{A}\pi^{B}+ \pi_{A}\omega^{B}\Bigr),\\
S(\omega,p){_{A'}}{^{B'}}&=&
-\frac{1}{2}\Bigl(
\bar \omega_{A'}\bar \pi^{B'}+ 
\bar \pi_{A'}\bar \omega^{B'}\Bigr),
\ee
and
\be
S(\omega,p){_{A}}{^{B}}\omega_{B}&=&
\frac{1}{2}\,\omega_{A},\\
S(\omega,p){_{A'}}{^{B'}}\bar \omega_{B'}&=&
-\frac{1}{2}\,\bar \omega_{A'},\\
S(\omega,p){_{A}}{^{B}}\pi_{B}&=&
-\frac{1}{2}\,\pi_{A},\\
S(\omega,p){_{A'}}{^{B'}}\bar \pi_{B'}&=&
\frac{1}{2}\,\bar \pi_{A'}.
\ee

\subsection{Eigenvectors of $S(\omega,p)_\alpha{^\beta}$}

Let 
\be
\pi_{\alpha}
&=&
-2\left(
\begin{array}{c}
\pi_A\\
\bar \pi_{A'}
\end{array}
\right)\label{38}
\ee
where $\pi_A=\pi_A(p)$ is the spinor appearing in the 
decomposition (\ref{p})
of a (massive or massless) 4-momentum $p$. 
The eigenvectors of the P-L vector projections in the
corresponding null
$\omega^a=\omega^a(p)$ 
direction can be defined in terms of $\pi_{\alpha}$:
\be
{\Omega}_\pm^{(+)}(\omega,p)_\alpha&=&
{\Pi}^{(+)}_{\pm}(\omega,p){_{\alpha}}{^{\beta}}
\pi_\beta
=
\left(
\begin{array}{c}
\pm\sqrt{\frac{p\cdot p}{2}}
\omega{_{A}}\label{I.40}\\
- \bar \pi{_{A'}}
\end{array}
\right)\\
{\Omega}_\pm^{(-)}(\omega,p)_\alpha&=&
{\Pi}^{(-)}_{\pm}(\omega,p){_{\alpha}}{^{\beta}}
\pi_\beta
=
\left(
\begin{array}{c}
- \pi{_{A}}\\
\mp\sqrt{\frac{p\cdot p}{2}}
\bar \omega{_{A'}}
\end{array}
\right).\label{I.41}
\ee
Notice that the massless limit $p\cdot p\to 0$, $p\neq 0$,  can be easily
performed for ${\Omega}_\pm^{(\pm)}(\omega,p)_\alpha$ 
whereas this would not be possible with the projectors
(\ref{30}), (\ref{31}) themselves. (We exclude the origin $p=0$
for which the whole construction breaks down since no spin-frame
satisfying our conditions exists at this point.) 
So the transvection with 
(\ref{38}) removes the parts which are singular in the massless
limit. This behavior is typical of the formalism developed
in this paper.

\section{Algebra of spin-energy projectors ($m\neq 0$)}
\label{Sec.III}

Formulas that look artificial and complicated at the bispinor
level finally simplify if one switches to unitary
representations. In order to do so we have to better control the
algebraic and transformation properties of the projectors.

\subsubsection{A few useful identities}

In this section we assume $p\cdot p\neq 0$. 
We shall derive several useful general formulas which, when
suitably transvected, will be directly related to the unitary
representations we are interested in, also for 
$p\cdot p\to 0$. Define
\be
{T}{^{\alpha}}{^{\beta'}}=
\left(
\begin{array}{c}
T{^{A}}{^{B'}}\\
0\\
0\\
T{^{B}}{^{A'}}
\end{array}
\right),
\ee
where $T^a$ is an arbitrary world-vector.
Below I list without proof a few useful identities satisfied by
projectors projecting on given signs of energy or spin
($m=\sqrt{p\cdot p}$):
\be
{T}{^{\alpha}}{^{\gamma'}}
{\Pi}_{\pm}(p){_\alpha}{^\beta}\bar
{\Pi}_{\pm}(p){_{\gamma'}}{^{\delta'}}&=&
\frac{\sqrt{2}}{4m}(T\cdot p)
\left(
\begin{array}{c}
\frac{\sqrt{2}}{m}p{^{B}}{^{D'}}
\\
\mp \ve^{BD}
\\
\pm \ve^{B'D'}
\\
\frac{\sqrt{2}}{m}p{^{D}}{^{B'}}
\end{array}
\right)\label{6}
\ee
\be
{T}{^{\alpha}}{^{\gamma'}}
{\Pi}_{\pm}(p){_\alpha}{^\beta}\bar
{\Pi}_{\mp}(p){_{\gamma'}}{^{\delta'}}&=&
\frac{1}{4}
\left(
\begin{array}{c}
2T{^{B}}{^{D'}}
-
\frac{2}{m^2}
(T\cdot p)
p{^{B}}{^{D'}}
\\
\mp \frac{\sqrt{2}}{m}\Bigl(
T{^{B}}{^{X'}}
 p{^{D}}{_{X'}}
+
T{^{D}}{^{X'}}
p{^{B}}{_{X'}}\Bigr)
\\
\pm \frac{\sqrt{2}}{m}
\Bigl(
T{^{X}}{^{B'}}
p{_{X}}{^{D'}}
+
T{^{X}}{^{D'}}
 p{_{X}}{^{B'}}
\Bigr)
\\
-\frac{2}{m^2}
(T\cdot p)
p{^{D}}{^{B'}}
+
2T{^{D}}{^{B'}}
\end{array}
\right).\label{17}
\ee
(\ref{6}) and (\ref{17}) imply
\be
{T}{^{\alpha}}{^{\gamma'}}
\Bigl(
{\Pi}_{+}(p){_\alpha}{^\beta}\bar
{\Pi}_{+}(p){_{\gamma'}}{^{\delta'}}
+
{\Pi}_{-}(p){_\alpha}{^\beta}\bar
{\Pi}_{-}(p){_{\gamma'}}{^{\delta'}}
\Bigr)
&=&
\frac{1}{m^2}
(T\cdot p)
{p}{^{\beta}}{^{\delta'}}\label{19}
\\
{T}{^{\alpha}}{^{\gamma'}}
\Bigl(
{\Pi}_{+}(p){_\alpha}{^\beta}\bar
{\Pi}_{-}(p){_{\gamma'}}{^{\delta'}}
+
{\Pi}_{-}(p){_\alpha}{^\beta}\bar
{\Pi}_{+}(p){_{\gamma'}}{^{\delta'}}
\Bigr)
&=&
{T}{^{\beta}}{^{\delta'}}
-
\frac{1}{m^2}
(T\cdot p)
{p}{^{\beta}}{^{\delta'}}.\label{20}
\ee
(\ref{19}) plays a role of a non-orthogonal ``resolution
of $p$" and is essential for the formalism developed below. 
Notice that for $T\cdot p\neq 0$, $p\cdot p\neq 0$  we can always write 
\be
\frac{1}{
p\cdot p}{p}{^{\alpha}}{^{\alpha'}}=
\frac{1}{
T\cdot p}
{T}{^{\beta}}{^{\beta'}}
\Bigl(
{\Pi}_{+}(p){_\beta}{^\alpha}\bar
{\Pi}_{+}(p){_{\beta'}}{^{\alpha'}}
+
{\Pi}_{-}(p){_\beta}{^\alpha}\bar
{\Pi}_{-}(p){_{\beta'}}{^{\alpha'}}
\Bigr)\label{21}
\ee
even though the ``off-diagonal" terms given by (\ref{20}) vanish
only for $T^a=p^a$. The RHS of Eq.~(\ref{21}) will be used in
definitions of positive-definite scalar products in momentum space. 

The operators projecting on eigenvectors of
$S(t,p)_\alpha{^\beta}$ are
\be
{\Pi}^{(\pm)}(t,p){_\alpha}{^\beta}
&=&
\frac{1}{2\lambda^{(\pm)}(t,p)}
\left(
\begin{array}{c}
\lambda^{(\pm)}(t,p)\varepsilon{_A}{^B} +
t{_{AX'}}p^{BX'}
-
p{_{A}}{_{X'}}t{^{B}}{^{X'}}
\\
0\\
0\\
\lambda^{(\pm)}(t,p)\varepsilon{_{A'}}{^{B'}} 
- t{_{XA'}}p^{XB'}
+
p{_{X}}{_{A'}}t{^{X}}{^{B'}}
\end{array}
\right).
\ee
They satisfy
\be
{}&{}&{T}{^{\alpha}}{^{\gamma'}}
{\Pi}_{\pm}(p){_\alpha}{^\beta}\bar
{\Pi}_{\pm}(p){_{\gamma'}}{^{\delta'}}
{\Pi}^{(\pm)}(t,p){_{\beta}}{^{\mu}}
\bar {\Pi}^{(\pm)}(u,p){_{\delta'}}{^{\nu'}}
:=
(T\cdot p)
{\Pi}_\pm^{(\pm\pm)}(t,u,p)^{\mu}{^{\nu'}}
=
\frac{T\cdot p}{8m^2\lambda(t,p)\lambda(u,p)}
\times
\nonumber\\
&{}&
\left(
\begin{array}{l}
\Bigl[
\Bigl(
\lambda^{(\pm)}(t,p)
+
t\cdot p
\Bigr)
\Bigl(
\lambda^{(\pm)}(u,p) 
+
u\cdot p
\Bigr)
-m^2\,t\cdot u
\Bigr]
p{^{M}}{^{N'}}
-
m^2
\Bigl(
\lambda^{(\pm)}(t,p)
u{^{M}}{^{N'}}
+
\lambda^{(\pm)}(u,p)t{^{MN'}}
+
i\,
e^{MN'abc}p_at_bu_c
\Bigr)\\
\mp
\frac{m}{\sqrt{2}}
\Bigl[
\Bigl(
\lambda^{(\pm)}(t,p)
\lambda^{(\pm)}(u,p)
+
(p\cdot t)(p\cdot u)
-
m^2\,t\cdot u
\Bigr)
\varepsilon{^M}{^N}
-
t{_{BX'}}u{^{B}}{_{Y'}}
p^{(M|X'}p^{|N)Y'}
-
\frac{3m^2}{2}t^{(M}{_{X'}}u{^{N)}}{^{X'}}+
\\
\hfill
+
\Bigl(
2\lambda^{(\pm)}(t,p)
+
p\cdot t
\Bigr)
p{^{(M}}{_{X'}}u{^{N)}}{^{X'}}
+
\Bigl(
2\lambda^{(\pm)}(u,p)
-
p\cdot u
\Bigr)
p^{(M}{_{X'}}
t{^{N)}}{^{X'}}\Bigr]
\\
\pm
\frac{m}{\sqrt{2}}
\Bigl[
\Bigl(
\lambda^{(\mp)}(t,p)
\lambda^{(\mp)}(u,p)
+
(p\cdot t)(p\cdot u)
-
m^2\,t\cdot u
\Bigr)
\varepsilon{^{M'}}{^{N'}}
-
t{_{XB'}}u{_{Y}}{^{B'}}
p^{X(M'|}p^{Y|N')}
-
\frac{3m^2}{2}t{_{X}}^{(M'|}u{^{X}}{^{|N')}}+
\\
\hfill
+
\Bigl(
2\lambda^{(\mp)}(t,p)
+
p\cdot t
\Bigr)
p{_{X}}{^{(M'|}}u{^{X}}{^{|N')}}
+
\Bigl(
2\lambda^{(\mp)}(u,p)
-
p\cdot u
\Bigr)
p{_{X}}^{(M'|}
t{^{X}}{^{|N')}}\Bigr]
\\
\Bigl[
\Bigl(
\lambda^{(\mp)}(t,p)
+
t\cdot p
\Bigr)
\Bigl(
\lambda^{(\mp)}(u,p) 
+
u\cdot p
\Bigr)
-m^2\,t\cdot u
\Bigr]
p{^{N}}{^{M'}}
-
m^2
\Bigl(
\lambda^{(\mp)}(t,p)
u{^{N}}{^{M'}}
+
\lambda^{(\mp)}(u,p)t{^{NM'}}
-
i\,
e^{NM'abc}p_at_bu_c
\Bigr)
\end{array}
\right)\nonumber\\
\label{15}
\ee
\be
{}&{}&
{T}{^{\alpha}}{^{\gamma'}}
{\Pi}_{\pm}(p){_\alpha}{^\beta}\bar
{\Pi}_{\pm}(p){_{\gamma'}}{^{\delta'}}
{\Pi}^{(\pm)}(t,p){_{\beta}}{^{\mu}}
\bar {\Pi}^{(\mp)}(u,p){_{\delta'}}{^{\nu'}}
:=
(T\cdot p)
{\Pi}_\pm^{(\pm\mp)}(t,u,p)^{\mu}{^{\nu'}}=
\frac{-T\cdot p}{8m^2\lambda(t,p)\lambda(u,p)}\times
\nonumber\\
&{}&
\left(
\begin{array}{l}
\Bigl[
\Bigl(
\lambda^{(\pm)}(t,p)
+
t\cdot p
\Bigr)
\Bigl(
\lambda^{(\mp)}(u,p) 
+
u\cdot p
\Bigr)
-m^2\,t\cdot u
\Bigr]
p{^{M}}{^{N'}}
-
m^2
\Bigl(
\lambda^{(\pm)}(t,p)
u{^{M}}{^{N'}}
+
\lambda^{(\mp)}(u,p)t{^{MN'}}
+
i\,
e^{MN'abc}p_at_bu_c
\Bigr)\\
\mp
\frac{m}{\sqrt{2}}
\Bigl[
\Bigl(
\lambda^{(\pm)}(t,p)
\lambda^{(\mp)}(u,p)
+
(p\cdot t)(p\cdot u)
-
m^2\,t\cdot u
\Bigr)
\varepsilon{^M}{^N}
-
t{_{BX'}}u{^{B}}{_{Y'}}
p^{(M|X'}p^{|N)Y'}
-
\frac{3m^2}{2}t^{(M}{_{X'}}u{^{N)}}{^{X'}}+
\\
\hfill
+
\Bigl(
2\lambda^{(\pm)}(t,p)
+
p\cdot t
\Bigr)
p{^{(M}}{_{X'}}u{^{N)}}{^{X'}}
+
\Bigl(
2\lambda^{(\mp)}(u,p)
-
p\cdot u
\Bigr)
p^{(M}{_{X'}}
t{^{N)}}{^{X'}}\Bigr]
\\
\pm
\frac{m}{\sqrt{2}}
\Bigl[
\Bigl(
\lambda^{(\mp)}(t,p)
\lambda^{(\pm)}(u,p)
+
(p\cdot t)(p\cdot u)
-
m^2\,t\cdot u
\Bigr)
\varepsilon{^{M'}}{^{N'}}
-
t{_{XB'}}u{_{Y}}{^{B'}}
p^{X(M'|}p^{Y|N')}
-
\frac{3m^2}{2}t{_{X}}^{(M'|}u{^{X}}{^{|N')}}+
\\
\hfill
+
\Bigl(
2\lambda^{(\mp)}(t,p)
+
p\cdot t
\Bigr)
p{_{X}}{^{(M'|}}u{^{X}}{^{|N')}}
+
\Bigl(
2\lambda^{(\pm)}(u,p)
-
p\cdot u
\Bigr)
p{_{X}}^{(M'|}
t{^{X}}{^{|N')}}\Bigr]
\\
\Bigl[
\Bigl(
\lambda^{(\mp)}(t,p)
+
t\cdot p
\Bigr)
\Bigl(
\lambda^{(\pm)}(u,p) 
+
u\cdot p
\Bigr)
-m^2\,t\cdot u
\Bigr]
p{^{N}}{^{M'}}
-
m^2
\Bigl(
\lambda^{(\mp)}(t,p)
u{^{N}}{^{M'}}
+
\lambda^{(\pm)}(u,p)t{^{NM'}}
-
i\,
e^{NM'abc}p_at_bu_c
\Bigr)
\end{array}
\right)\nonumber\\\label{16}
\ee
where $\lambda(t,p)=|\lambda^{(\pm)}(t,p)|$, etc. 
Although (\ref{15}), (\ref{16}) may appear somewhat complicated,
their generality will finally help us to simplify the whole formalism. 

\subsubsection{$SL(2,C)$ active transformations of the projectors}

The projectors transform under {\it active\/} $SL(2,C)$ 
as follows
\be
{S}{_{\alpha}}{^{\gamma}}
{S}{_{\beta}}{^{\delta}}
{\Pi}_{\pm}(p)_{\gamma\delta}
&=&{\Pi}_{\pm}(S p){_\alpha}{_\beta}\\
{S}{_{\alpha}}{^{\gamma}}
{S}{_{\beta}}{^{\delta}}
{\Pi}^{(\pm)}(t,p)_{\gamma\delta}
&=&{\Pi}^{(\pm)}(S t,S p){_\alpha}{_\beta}
\ee
or equivalently
\be
{S}{_{\alpha}}{^{\beta}}
{\Pi}_{\pm}(S^{-1}p){_\beta}{^\gamma}{S}^{-1}{_{\gamma}}{^{\delta}}
&=&{\Pi}_{\pm}(p){_\alpha}{^\delta}\label{I.53}\\
{S}{_{\alpha}}{^{\beta}}
{\Pi}^{(\pm)}(S^{-1}t,
S^{-1}p){_\beta}{^\gamma}{S}^{-1}{_{\gamma}}{^{\delta}}
&=&{\Pi}^{(\pm)}(t,p){_\alpha}{^\delta}\label{I.54}
\ee
where 
\be
{S}_{\alpha}{^{\beta}}=
\left(
\begin{array}{c}
S_{A}{^{B}}\\
0\\
0\\
S_{A'}{^{B'}}
\end{array}
\right).
\ee
Here $(Sp)_a=S_a{^b}p_b$, and $S_a{^b}$, ${S}_{\alpha}{^{\beta}}$, 
${S}_{A}{^{B}}$, and ${S}_{A'}{^{B'}}$ denote, respectively, the
representations $(1/2,1/2)$, $(1/2,0)\oplus(0,1/2)$, 
$(1/2,0)$,  and $(0,1/2)$ of $S\in SL(2,C)$. 
Spinor transformations of upper- and lower-index spinors are
assumed in the form
\be
S\phi_{A}&=&S{_{A}}{^{B}}\phi_{B}\label{cov},\\
S\phi^{A}&=&\phi^{B}S^{-1}{_{B}}{^{A}}=-S{^{A}}{_{B}}\phi^{B}.
\label{con}
\ee
Analogous transformations hold for primed spinors.
The convention differs slightly from this used in 
\cite{PR} (cf. Eq.~(3.6.1)) but seems more consistent.
Those {\it active nonunitary\/} transformations will be shown to
generate the {\it passive unitary\/} transformations which form
the unitary represenations of the Poincar\'e group we are
searching for.

\section{Wave equations associated with the projectors}
\label{Sec.IV}

It is clear from what has been written above what is the
relation of the projectors to the ``sign of spin". To understand
their relation to the ``sign of energy" we have to discuss their
relation to Bargmann-Wigner equations. A brief
analysis of the equations will also help to naturally clasify
representations with respect to the signs of energy and {\it mass\/}.

\subsection{Off-shell equations}

Define
\be
{D}{_\alpha}{^\beta}=
\frac{1}{\sqrt{2}}
\left(
\begin{array}{c}
\frac{i}{\sqrt{2}}
\varepsilon{_A}{^B}\partial_s \\
i\nabla{_{A}}{^{B'}}\\
-i\nabla{^{B}}{_{A'}} \\
\frac{i}{\sqrt{2}}
\varepsilon{_{A'}}{^{B'}}\partial_s
\end{array}
\right)
\ee
satisfying
\be
{D}{_\alpha}{^\beta}e^{i s\sqrt{p\cdot p}\mp i p\cdot x}
&=&
-\sqrt{p\cdot p}\,
{\Pi}_\mp (p){_\alpha}{^\beta}
e^{i s\sqrt{p\cdot p}\mp i p\cdot x},\\
{D}{_\alpha}{^\beta}e^{-i s\sqrt{p\cdot p}\mp i p\cdot x}
&=&
\sqrt{p\cdot p}\,
{\Pi}_\pm (p){_\alpha}{^\beta}
e^{-i s\sqrt{p\cdot p}\mp i p\cdot x}.
\ee
The off-shell Bargamann-Wigner equation
\be
{D}{_{\alpha_k}}{^{\beta_k}}\psi(x,s)_
{\beta_1\dots\beta_k\dots\beta_n}&=&0
\ee
has ``plane-wave" solutions of the form
\be
\psi(p,x,s)_{\alpha_1\dots\alpha_n}
&=&
{\Pi}_+ (p){_{\alpha_1}}{^{\beta_1}}
\dots
{\Pi}_+ (p){_{\alpha_n}}{^{\beta_n}}
\Bigl(
\psi_{+-}(p)_{\beta_1\dots\beta_n}
e^{-i s\sqrt{p\cdot p}+ i p\cdot x}
+
\psi_{-+}(p)_{\beta_1\dots\beta_n}
e^{i s\sqrt{p\cdot p}- i p\cdot x}
\Bigr)
\nonumber\\
{}&{}&
+
{\Pi}_- (p){_{\alpha_1}}{^{\beta_1}}
\dots
{\Pi}_- (p){_{\alpha_n}}{^{\beta_n}}
\Bigl(
\psi_{--}(p)_{\beta_1\dots\beta_n}
e^{i s\sqrt{p\cdot p}+ i p\cdot x}
+
\psi_{++}(p)_{\beta_1\dots\beta_n}
e^{-i s\sqrt{p\cdot p}- i p\cdot x}
\Bigr).\label{plane wave}
\ee
Let us note that the formula
\be
{\Pi}_\pm (p){_\alpha}{^\beta}=
{\Pi}_\mp (-p){_\alpha}{^\beta}
\ee
guarantees that nothing physical will be lost by assuming that
$p$ is {\it future pointing\/}. Throughout the rest of the paper
I therefore assume that
$\psi_{\cdot\,\cdot}(p)_{\beta_1\dots\beta_n}=0$ 
for $p$ pointing into the past.

The superpositions 
\be
\psi(x,s)_{\alpha_1\dots\alpha_n}=\frac{1}{(2\pi)^4}\int d^4p\,
\psi(p,x,s)_{\alpha_1\dots\alpha_n},\label{wave packet}
\ee
with arbitrary higher rank bispinors 
$\psi_{\cdot\,\cdot}(p)_{\beta_1\dots\beta_n}$, play a role of ``proper
time" wave packets in Minkowski space.

\subsection{On-shell equations}

Changing variables $(p_0,\bbox p)\to (\sqrt{p\cdot p},\bbox p)$
we obtain 
\be
{}&{}&
\frac{1}{(2\pi)^4}\int d^4p\,
\psi(p,x,s)_{\alpha_1\dots\alpha_n}=
\frac{1}{(2\pi)^4}\int d(p\cdot p)\int \frac{d^3p}
{2\sqrt{\bbox{p}^2 + p\cdot p}}
\psi(\sqrt{\bbox{p}^2 + p\cdot p}, \bbox
p,x,s)_{\alpha_1\dots\alpha_n}.\nonumber 
\ee
The on-shell bispinors satisfy
$\psi_{\cdot\,\cdot}(p)=2\pi\psi_{m^2{\cdot\,\cdot}}
(p)\mu(p\cdot p-m^2)$ where $\mu(p\cdot p-m^2)$ is some
distribution concentrated on the mass-$m$ hyperboloid. 
We will consider two cases: 
$\mu(p\cdot p-m^2)=\delta(p\cdot p-m^2)$ and 
$\mu(p\cdot p-m^2)=\sqrt{\delta(p\cdot p-m^2)}$.
For 
$\mu(p\cdot p-m^2)=\delta(p\cdot p-m^2)$ we obtain 
\be
\frac{1}{(2\pi)^4}\int d^4p\,
\psi(p,x,s)_{\alpha_1\dots\alpha_n}
=
\frac{1}{(2\pi)^3}\int \frac{d^3p}
{2|p_m^0|}
\psi_{m^2}(p_m,x,s)_{\alpha_1\dots\alpha_n}
\label{on-D}
\ee
where, by definition, $p_m=(\sqrt{\bbox{p}^2 + m^2}, \bbox p)$
is future-pointing. 
The combinations where $\psi_{m^2-\,\cdot}=0$ 
($\psi_{m^2+\,\cdot}=0$)
are solutions of
the positive(negative)-mass on-shell Bargmann-Wigner equation. 
There exist also other possibilities of transition to an on-shell
formalism \cite{Almond1,Almond2}.

\section{Positive-definite scalar products in momentum space}
\label{Sec.V}

There exists a class of equivalent positive-definite scalar products that
can be used to probabistically interpret the Fourier components
of Bargmann-Wigner fields. They can be simplified at the unitary
representation level.  

\subsection{Off-shell products}

Consider some (in general $p$-dependent) $T^a$ satisfying, for
the time being, $T\cdot p\neq 0$. 
For two off-shell functions $\psi(x,s)_{\alpha_1\dots\alpha_n}$, 
$\phi(x,s)_{\alpha_1\dots\alpha_n}$ the scalar product is
\be
\langle\psi,\phi\rangle 
&=&
\frac{1}{(2\pi)^4}\int d^4p\,
\frac{1}{(T\cdot p)^n}
{T}^{\alpha_1\alpha'_1}\dots  {T}^{\alpha_n\alpha'_n}
\times\nonumber\\
&\pp =&
\Biggl[
{\Pi}_+ (p){_{\alpha_1}}{^{\beta_1}}
\bar {\Pi}_+ (p){_{\alpha'_1}}{^{\beta'_1}}
\dots
{\Pi}_+(p){_{\alpha_n}}{^{\beta_n}}
\bar {\Pi}_+(p){_{\alpha'_n}}{^{\beta'_n}}
\Bigl(
\bar \psi_{+-}(p)_{\beta'_1\dots\beta'_n}
\phi_{+-}(p)_{\beta_1\dots\beta_n}
+
\bar \psi_{-+}(p)_{\beta'_1\dots\beta'_n}
\phi_{-+}(p)_{\beta_1\dots\beta_n}
\Bigr)\nonumber\\
&\pp =&
+
{\Pi}_- (p){_{\alpha_1}}{^{\beta_1}}
\bar {\Pi}_- (p){_{\alpha'_1}}{^{\beta'_1}}
\dots
{\Pi}_-(p){_{\alpha_n}}{^{\beta_n}}
\bar {\Pi}_-(p){_{\alpha'_n}}{^{\beta'_n}}
\Bigl(
\bar \psi_{--}(p)_{\beta'_1\dots\beta'_n}
\phi_{--}(p)_{\beta_1\dots\beta_n}
+
\bar \psi_{++}(p)_{\beta'_1\dots\beta'_n}
\phi_{++}(p)_{\beta_1\dots\beta_n}
\Bigr)
\Biggr].\nonumber\\
\label{off-sp}
\ee

\subsection{On-shell products}

The on-shell scalar product of two bispinors 
(\ref{on-D}) will be denoted by
\be
\langle\psi,\phi\rangle_{m^2}\label{on-sp}
\ee
and is obtained from (\ref{off-sp}) by 
$p\to p_m$, $\psi_{{\cdot\,\cdot}}\to\psi_{m^2{\cdot\,\cdot}}$, and 
$$
\frac{1}{(2\pi)^4}\int d^4p
\to
\frac{1}{(2\pi)^3}\int \frac{d^3p}
{2|p_m^0|}.
$$
It can be also thought of as an on-shell limit of the off-shell
formula (\ref{off-sp}) provided the limiting bispinors are 
\be
\psi_{\cdot\,\cdot}(p)=2\pi\psi_{m^2{\cdot\,\cdot}}(p)
\sqrt{\delta(p\cdot p-m^2)}.
\ee
Putting $T=p_m$, $m\neq 0$ or using the resolution of $p$
formula (\ref{19}), we obtain the form used by
Woodhouse in \cite{W}. The well known Bargmann-Wigner form 
\cite{BW} is
obtained if $T$ is a time direction and $T\cdot p_m=p_m^0$. 
The use of projectors makes it similar to the form used by
Kaiser \cite{K2} in his construction of electromagnetic
wavelets. Eq.~(\ref{19}) shows explicitly that the products
are $T$-independent.

\section{Momentum-space amplitudes}
\label{Sec.VI}

Momentum-space amplitudes have a direct probability
interpretation and have been extensively discussed in literature
\cite{M1,M2,M3,M4-1,M4-2,IBB3,IBBqed}. From the viewpoint of active $SL(2,C)$ 
transformations they are scalars. They become local (i.e.
$p$-dependent) $SU(2)$ spinors if passive transformations are
concerned. Typically they are represented as helicity
amplitudes. Here we find their general form and from this
perspective find their most convenient representation in terms of
$\omega$-spinors.

\subsection{Normalized spin-energy eigenvectors}
\label{normalization}

\subsubsection{Normalization factor}

The particular cases of (\ref{16}) and (\ref{15}) for $t=u$ 
\be
{\Pi}_\pm^{(\pm\mp)}(t,t,p)^{\mu}
{^{\nu'}}
&=&0\\
{\Pi}_\pm^{(\pm\pm)}(t,t,p)^{\mu}
{^{\nu'}}
&=&
\frac{1}{4m^2\lambda(t,p)}
\left(
\begin{array}{l}
\Bigl[
\lambda(t,p)
(\pm) 
t\cdot p
\Bigr]
p{^{M}}{^{N'}}
(\mp)
m^2
t{^{M}}{^{N'}}
\\
\mp
\frac{m}{\sqrt{2}}
\Bigl[
\lambda(t,p)
\varepsilon{^M}{^N}
(\pm)
2
p{^{(M}}{_{X'}}t{^{N)}}{^{X'}}
\Bigr]
\\
\pm
\frac{m}{\sqrt{2}}
\Bigl[
\lambda(t,p)
\varepsilon{^{M'}}{^{N'}}
(\mp)
2
p{_{X}}{^{(M'|}}t{^{X}}{^{|N')}}
\Bigr]
\\
\Bigl[
\lambda(t,p)
(\mp)
t\cdot p
\Bigr]
p{^{N}}{^{M'}}
(\pm)
m^2
t{^{N}}{^{M'}}
\end{array}
\right)
\ee
can be used to find the correct normalization factor for 
the spin-energy eigenvectors for an arbitrary $t$:
\be
{\Pi}_\pm^{(\pm\pm)}(t,t,p)^{\mu}{^{\nu'}}
{\Omega}_\pm^{(\pm)}(\omega,p)_\mu
\bar {\Omega}_\pm^{(\pm)}(\omega,p)_{\nu'}
&=&
\frac{1}{2\lambda(t,p)}
\Biggl\{
\lambda(t,p)
+
t\cdot\Bigl(
\pi
-
\frac{m^2}{2}
\omega
\Bigr)
\Biggr\}=:N(t,\pi,\omega,m)^2.
\ee
The mentioned similarity between the massless and massive cases for
$t=\omega$, characteristic of the null formalism, can be seen
also in the formula
\be
N(t,\pi,\omega,m=0)=N(\omega,\pi,\omega,m\neq 0)=1.
\ee

\subsubsection{Eigenvectors}

The normalized eigenvectors are 
\be
{\Omega}_\pm^{(\pm)}(t,p)_\alpha=
N(t,\pi,\omega,m)^{-1}
{\Pi}^{(\pm)}_{\pm}(t,p){_{\alpha}}{^{\beta}}
{\Omega}_\pm^{(\pm)}(\omega,p)_\beta.
\ee
They satisfy
\be
\frac{1}{T\cdot p}
T^{\alpha\alpha'}
{\Omega}_\pm^{(\pm)}(t,p)_\alpha
\bar {\Omega}_\pm^{(\pm)}(t,p)_{\alpha'}&=&1\\
\frac{1}{T\cdot p}
T^{\alpha\alpha'}
{\Omega}_\pm^{(\pm)}(t,p)_\alpha
\bar {\Omega}_\pm^{(\mp)}(t,p)_{\alpha'}&=&0,
\ee
which hold also for $p\cdot p=0$ and the 0/0-type limit $T\to p$.
The fact that neither orthogonality nor normalization depend on
$T$ can be used to simplify the formalism by choosing
$T=\omega$. With this choice we have
\be
\omega^{\alpha\alpha'}
{\Omega}_\pm^{(\pm)}(t,p)_\alpha
\bar {\Omega}_\pm^{(\pm)}(t,p)_{\alpha'}&=&1.
\ee
The explicit forms of the eigenvectors are
\be
\Omega_\pm^{(+)}(t,p)_\alpha
&=&
\Bigl[8\lambda(t,p)
\bigl(
\lambda(t,p)
+
t\cdot p
-
(p\cdot p)
(t\cdot\omega)
\bigr)\Bigr]^{-1/2}\nonumber\\
&\pp =&\times
\left(
\begin{array}{c}
\pm\sqrt{\frac{p\cdot p}{2}}
\Bigl(
(2\lambda(t,p)-t\cdot p) \omega{_{A}}
+\omega{_{C}}t{^C}{_{X'}}p{_{A}}{^{X'}} +3 t{_{AX'}}\bar \pi{^{X'}}
\Bigr)
\\
-
(2\lambda(t,p)+t\cdot p) 
\bar \pi{_{A'}}
+ \bar \pi{_{C'}}t{_X}{^{C'}}p{^{X}}{_{A'}} 
- 3\frac{m^2}{2}t{_{XA'}}\omega{^{X}}
\end{array}
\right)\\
&=&
\left(
\begin{array}{c}
{\Omega}_\pm^{(+)}(t,p)_A\\
{\Omega}_\pm^{(+)}(t,p)_{A'}
\end{array}
\right)\\
\Omega_\pm^{(-)}(t,p)_\alpha
&=&
\Bigl[8\lambda(t,p)
\bigl(
\lambda(t,p)
+
t\cdot p
-
(p\cdot p)
(t\cdot\omega)
\bigr)\Bigr]^{-1/2}\nonumber\\
&\pp =&\times
\left(
\begin{array}{c}
-
(2\lambda(t,p)+t\cdot p) 
\pi{_{A}}
+
\pi{_{C}}t{^C}{_{X'}}p{_{A}}{^{X'}} 
-
3\frac{m^2}{2}t{_{AX'}}\bar \omega{^{X'}}
\\
\mp\sqrt{\frac{p\cdot p}{2}}
\Bigl(
(2\lambda(t,p)-t\cdot p) 
\bar \omega{_{A'}}
+
\bar \omega{_{C'}}t{_X}{^{C'}}p{^{X}}{_{A'}} 
+
3t{_{XA'}}\pi{^{X}}\Bigr) 
\end{array}
\right)\\
&=&
\left(
\begin{array}{c}
{\Omega}_\pm^{(-)}(t,p)_A\\
{\Omega}_\pm^{(-)}(t,p)_{A'}
\end{array}
\right)
=
\left(
\begin{array}{c}
\bar {\Omega}_\pm^{(+)}(t,p)_A\\
-\bar {\Omega}_\pm^{(+)}(t,p)_{A'}
\end{array}
\right)\label{86}
\ee
The massless limit is easy to perform:
\be
{\Omega}_\pm^{(+)}(t,p)_\alpha&\to&
\left(
\begin{array}{c}
0
\\
-
\bar \pi{_{A'}}
\end{array}
\right),\\
{\Omega}_\pm^{(-)}(t,p)_\alpha&\to&
\left(
\begin{array}{c}
-
\pi{_{A}}
\\
0
\end{array}
\right).
\ee
These eigenvectors do not depend on $t$ which is what one should
expect since components of the P-L vector commute in this limit.

\subsection{Off-shell amplitudes}

\subsubsection{Dirac bispinors}

We shall first concentrate on the Dirac bispinors (i.e.
bispinors of rank 1). The wave functions associated with them
play a role analogous to ordinary 2-spinors. Consider the
bispinor (\ref{plane wave}) for $n=1$
\be
\psi(p,x,s)_\alpha&=&
{\Pi}_+ (p){_\alpha}{^\beta}
\Bigl(
\psi_{+-}(p)_\beta
e^{-i s\sqrt{p\cdot p}+ i p\cdot x}
+
\psi_{-+}(p)_\beta
e^{i s\sqrt{p\cdot p}- i p\cdot x}
\Bigr)
\nonumber\\
{}&{}&
+
{\Pi}_- (p){_\alpha}{^\beta}
\Bigl(
\psi_{--}(p)_\beta
e^{i s\sqrt{p\cdot p}+ i p\cdot x}
+
\psi_{++}(p)_\beta
e^{-i s\sqrt{p\cdot p}- i p\cdot x}
\Bigr)\nonumber\\
&=&
\sum_{(\pm)}
\Bigl(
{\Omega}_+^{(\pm)}(t,p)_\alpha
f_+(t,p)^{+}_{(\pm)}
e^{-i s\sqrt{p\cdot p}+ i p\cdot x}
+
{\Omega}_+^{(\pm)}(t,p)_\alpha
f_+(t,p)^{-}_{(\pm)}
e^{i s\sqrt{p\cdot p}- i p\cdot x}
\nonumber\\
{}&{}&
\pp {\sum_{(\pm)}
\Bigl(}
+
{\Omega}_-^{(\pm)}(t,p)_\alpha
f_-(t,p)^{-}_{(\pm)}
e^{i s\sqrt{p\cdot p}+ i p\cdot x}
+
{\Omega}_-^{(\pm)}(t,p)_\alpha
f_-(t,p)^{+}_{(\pm)}
e^{-i s\sqrt{p\cdot p}- i p\cdot x}
\Bigr)\nonumber\\
&=&
{\Omega}_+^{\cal A}(t,p)_\alpha
\Bigl(
f_+(t,p)^{+}_{\cal A}
e^{-i s\sqrt{p\cdot p}+ i p\cdot x}
+
f_+(t,p)^{-}_{\cal A}
e^{i s\sqrt{p\cdot p}- i p\cdot x}
\Bigr)
\nonumber\\
{}&{}&
+
{\Omega}_-^{\cal A}(t,p)_\alpha
\Bigl(
f_-(t,p)^{-}_{\cal A}
e^{i s\sqrt{p\cdot p}+ i p\cdot x}
+
f_-(t,p)^{+}_{\cal A}
e^{-i s\sqrt{p\cdot p}- i p\cdot x}
\Bigr)
\ee
where the calligraphic indices $\cal A$ equal $(\pm)$ and a
summation convention has been applied. The
amplitudes $f_\pm(t,p)^{\dots}_{\cal A}$ are  {\it
scalars\/} if {\it active\/} $SL(2,C)$ transformations are
concerned. An active Poincar\'e transformation acts on
(\ref{wave packet})
as follows
\be
{}&{}&
{\cal P}(S,a)\psi(x,s)_\alpha=S_{\alpha}{^{\beta}}
\psi\bigl(S^{-1}(x-a),s\bigr)_\beta\\
&{}&=
\frac{1}{(2\pi)^4}
\int d^4p\,\nonumber\\
&{}&\pp =
\Biggl[
S_{\alpha}{^{\beta}}
{\Omega}_+^{\cal A}(t[S^{-1}p],S^{-1}p)_\beta
\Bigl(
e^{-ip\cdot a}
f_+(t[S^{-1}p],S^{-1}p)^{+}_{\cal A}
e^{-i s\sqrt{p\cdot p}+ i p\cdot x}
+
e^{ip\cdot a}
f_+(t[S^{-1}p],S^{-1}p)^{-}_{\cal A}
e^{i s\sqrt{p\cdot p}- i p\cdot x}
\Bigr)
\nonumber\\
{}&{}&\pp =
+
S_{\alpha}{^{\beta}}
{\Omega}_-^{\cal A}(t[S^{-1}p],S^{-1}p)_\beta
\Bigl(
e^{-ip\cdot a}
f_-(t[S^{-1}p],S^{-1}p)^{-}_{\cal A}
e^{i s\sqrt{p\cdot p}+ i p\cdot x}
+
e^{ip\cdot a}
f_-(t[S^{-1}p],S^{-1}p)^{+}_{\cal A}
e^{-i s\sqrt{p\cdot p}- i p\cdot x}
\Bigr)
\Biggr]\label{t[p]}.
\ee
In (\ref{t[p]}) we have taken care of the fact that $t$ is in
general $p$-dependent.
The four types of amplitudes lead to the four classes of
momentum space representations:
\be
\psi_+(p)_\alpha^{+}={\Omega}_+^{\cal A}(t,p)_\alpha
f_+(t,p)^{+}_{\cal A}
&\to&
e^{-ip\cdot a}
S_{\alpha}{^{\beta}}{\Omega}_+^{\cal A}(t[S^{-1}p],S^{-1}p)_\beta
f_+(t[S^{-1}p],S^{-1}p)^{+}_{\cal A}\label{I.91}\\
\psi_+(p)_\alpha^{-}={\Omega}_+^{\cal A}(t,p)_\alpha
f_+(t,p)^{-}_{\cal A}
&\to&
e^{ip\cdot a}
S_{\alpha}{^{\beta}}{\Omega}_+^{\cal A}(t[S^{-1}p],S^{-1}p)_\beta
f_+(t[S^{-1}p],S^{-1}p)^{-}_{\cal A}\\
\psi_-(p)_\alpha^{-}={\Omega}_-^{\cal A}(t,p)_\alpha
f_-(t,p)^{-}_{\cal A}
&\to&
e^{-ip\cdot a}
S_{\alpha}{^{\beta}}{\Omega}_-^{\cal A}(t[S^{-1}p],S^{-1}p)_\beta
f_-(t[S^{-1}p],S^{-1}p)^{-}_{\cal A}\\
\psi_-(p)_\alpha^{+}={\Omega}_-^{\cal A}(t,p)_\alpha
f_-(t,p)^{+}_{\cal A}
&\to&
e^{ip\cdot a}
S_{\alpha}{^{\beta}}{\Omega}_-^{\cal A}(t[S^{-1}p],S^{-1}p)_\beta
f_-(t[S^{-1}p],S^{-1}p)^{+}_{\cal A}.\label{I.94}
\ee
Apparently there are only two types of transformations here, so
it may seem artificial to divide them into four classes. 
Later we shall see, however, 
that the above {\it active\/} and {\it nonunitary\/}
representations lead indeed to four different classes of {\it
passive unitary\/} transformations of the amplitudes. 
The amplitudes satisfy
\be
f_\pm(t,p)^{\dots}_{\cal A}=
\frac{1}{T\cdot p}T^{\alpha\alpha'}
\psi_\pm(p)_\alpha^{\dots}
\bar {\Omega}_\pm^{\cal A}(t,p)_{\alpha'}
=
\omega^{\alpha\alpha'}
\psi_\pm(p)_\alpha^{\dots}
\bar {\Omega}_\pm^{\cal A}(t,p)_{\alpha'}.\label{f(t,p)}
\ee
Of particular importance are the amplitudes
$f_\pm(\omega,p)^{\dots}_{\cal A}$ since
\be
\psi_\pm(p)_\alpha^{\dots}
&=&
{\Omega}_\pm^{(+)}(\omega,p)_\alpha
f_\pm(\omega,p)^{\dots}_{(+)}
+
{\Omega}_\pm^{(-)}(\omega,p)_\alpha
f_\pm(\omega,p)^{\dots}_{(-)}\nonumber\\
&=&
\left(
\begin{array}{c}
\pm\sqrt{\frac{p\cdot p}{2}}
\omega{_{A}}\\
- \bar \pi{_{A'}}
\end{array}
\right)f_\pm(\omega,p)^{\dots}_{(+)}
+
\left(
\begin{array}{c}
- \pi{_{A}}\\
\mp\sqrt{\frac{p\cdot p}{2}}
\bar \omega{_{A'}}
\end{array}
\right)
f_\pm(\omega,p)^{\dots}_{(-)}\label{I.96}
\ee
implies the following simple rule
\be
\omega^A
\psi_\pm(p)_A^{\dots}
&=&
f_\pm(\omega,p)^{\dots}_{(-)}
=:
f_\pm(\omega,p)^{\dots}_{\it 0}\label{BW1}\\
\bar \omega^{A'}
\psi_\pm(p)_{A'}^{\dots}
&=&
f_\pm(\omega,p)^{\dots}_{(+)}
=:
f_\pm(\omega,p)^{\dots}_{\it 1}.\label{BW2}
\ee
Formulas (\ref{BW1}), (\ref{BW1}) suggest that complex conjugation should
exchange {\it 0\/}'s  and {\it 1\/}'s:
\be
\overline{
f_{\dots}(\omega,p)^{\dots}_{\it 0}
}
&=&
\bar f_{\dots}(\omega,p)^{\dots}_{\it 1}\label{ccBW1}\\
\overline{
f_{\dots}(\omega,p)^{\dots}_{\it 1}
}
&=&
\bar f_{\dots}(\omega,p)^{\dots}_{\it 0}.\label{ccBW2}
\ee
The convention is consistent with (\ref{cc-of-S}).  Later we
shall see that complex conjugation exchanges also pluses and minuses.
The $\omega$-amplitudes play a distinguished role in the formalism
discussed in this paper. They will be shown to transform
{\it passively\/} as local $SU(2)$ spinors. We shall call them
the $\omega$-spinors. The general $t$-amplitudes
will be called the Bargmann-Wigner spinors (BW-spinors) or
simply the $t$-spinors.
The BW-spinor indices will be written in italic font to distinguish
them from the ordinary $SL(2,C)$ ones. 
The $\omega$-spinors appear implicitly in wave functions discussed
in the context of geometric quantization of spinor fields
\cite{W} where they are introduced purely formally as polarized
sections. The spin-frame decomposition used in \cite{W} is not
(\ref{p}) but 
\be
p^a=\frac{m}{\sqrt{2}}
\Bigl(
\pi^{A} \bar \pi^{A'}
+
\omega^{A} \bar \omega^{A'}
\Bigr),\label{Wp}
\ee
which is not very practical from our perspective since it does not
allow for a well defined $p\cdot p\to 0$ limit. 

Finally let us note that similar objects but in Minkowski four-{\it
position\/} representation are used to define spin-weighted
spherical harmonics \cite{PR}. 

\subsubsection{Equivalence between $\omega$-spinors and certain
$t$-spinors for spacelike or timelike $t$ and $m\neq 0$}

Standard spin eigenstates known from nonrelativistic quantum
mechanics correspond to spacelike $t$'s. 
Their relation with the
P-L vector has been discussed in detail in \cite{EPRB}.
The helicity formalism
is obtained by taking $t$ timelike.
In this context our choice of null $t=\omega$ is somewhat
counter-intuitive. Let us observe, however, that the eigenvalue
problem for the P-L vector possesses a kind of gauge freedom:
The eigenstates of $S(t,p)$ are unchanged by the transformation 
\be
t\to t+ \theta\, p.\label{gauge}
\ee
For $m\neq 0$ consider the spacelike 
\be
t=\omega -m^{-2}p\label{gauge1}
\ee
satisfying $t\cdot p=0$. 
The eigenvalues of $S(t,p)$ equal $\pm 1/2$ and 
$f_\pm(t,p)^{\dots}_{\cal A}=f_\pm(\omega,p)^{\dots}_{\cal A}$.
The orthogonality of $t$ and $p$ means that we consider a kind
of rest-frame eigenvalue problem for spin. Had we chosen
$\theta>0$ we would have obtained some sort of helicity
formalism and still the same amplitudes.

\subsubsection{Transition between general $t$-spinors and
$\omega$-spinors} 

Eqs.~(\ref{86}), (\ref{f(t,p)}), (\ref{BW1}) and (\ref{BW2}) imply 
\be
f_\pm(t,p)^{\dots}_{\cal A}=
{\cal W}_\pm(t,\omega,p)_{\cal A}{^{\cal B}}
f_\pm(\omega,p)^{\dots}_{\cal B},\label{calW}
\ee
which in terms of components reads
\be
\left(
\begin{array}{c}
f_\pm(t,p)^{\dots}_{\it 0}\\
f_\pm(t,p)^{\dots}_{\it 1}
\end{array}
\right)
&=&
\left(
\begin{array}{cc}
\bar \omega^{A'}
\bar {\Omega}_\pm^{(-)}(t,p)_{A'} &
\omega^{A}
\bar {\Omega}_\pm^{(-)}(t,p)_{A}\\
-\bar \omega^{A'}
{\Omega}_\pm^{(-)}(t,p)_{A'} &
\omega^{A}
{\Omega}_\pm^{(-)}(t,p)_{A}
\end{array}
\right)
\left(
\begin{array}{c}
f_\pm(\omega,p)^{\dots}_{\it 0}\\
f_\pm(\omega,p)^{\dots}_{\it 1}
\end{array}
\right).
\ee
${\cal W}_\pm(t,\omega,p)$ is unimodular
\be
\det {\cal W}_\pm(t,\omega,p)=
\omega^{\alpha\alpha'}
{\Omega}_\pm^{(-)}(t,p)_\alpha
\bar {\Omega}_\pm^{(-)}(t,p)_{\alpha'}
=1
\ee
and hence also unitary. From now on we shall concentrate
exclusively on $\omega$-spinors.

\subsubsection{$\ve$ and $\vs$ BW-spinors}

BW-spinors can be raised and lowered according to the standard
rules by 
\be
\ve_{\cal AB}=
\left(
\begin{array}{cc}
0 & 1 \\
-1 & 0 
\end{array}
\right)=\ve^{\cal AB}.\label{ve}
\ee
The fact that BW-spinors are local SU(2) spinors means that in
addition to $\ve$'s
there exists another invariant BW-spinor
\be
\varsigma^{\cal AB}=
\left(
\begin{array}{cc}
0 & 1 \\
1 & 0 
\end{array}
\right)=-\varsigma_{\cal AB}.\label{vs}
\ee
Unitarity of a unimodular ${\cal W}$ means 
\be
\vs_{\cal AB} &=&  
\bar {\cal W}{_{\cal A}}{^{\cal C}}\,
{\cal W}{_{\cal B}}{^{\cal D}}\,
\vs_{\cal CD}.
\ee
In terms of components:
\be
{\cal W}{_{\it 0}}{^{\it 0}}&=&
\overline{
{\cal W}{_{\it 1}}{^{\it 1}}}=
\bar {\cal W}{_{\it 0}}{^{\it 0}},\label{I.109}\\
{\cal W}{_{\it 1}}{^{\it 0}}&=&
-\,\overline{
{\cal W}{_{\it 0}}{^{\it 1}}}=-\,
\bar {\cal W}{_{\it 1}}{^{\it 0}}.\label{I.110}
\ee
If $f_{\cal A}$ is a BW-spinor then 
\be
\vs^{\cal AB}f_{\cal A}\bar f_{\cal B}=|f_{\it 0}|^2+|f_{\it 1}|^2.
\ee

\subsubsection{Higher-rank $\omega$-spinors}

Consider a Bargmann-Wigner rank-$n$ bispinor
$\psi(p,x,s)_{\alpha_1\dots\alpha_n}$. Its spin-energy
expansion is
\be
{}&{}&
\psi(p,x,s)_{\alpha_1\dots\alpha_n}\nonumber\\
&{}&={\Omega}_+^{{\cal A}_1}(t,p)_{\alpha_1}
\dots
{\Omega}_+^{{\cal A}_n}(t,p)_{\alpha_n}
\Bigl(
f_+(t,p)^{+}_{{\cal A}_1\dots{\cal A}_n}
e^{-i s\sqrt{p\cdot p}+ i p\cdot x}
+
f_+(t,p)^{-}_{{\cal A}_1\dots{\cal A}_n}
e^{i s\sqrt{p\cdot p}- i p\cdot x}
\Bigr)
\nonumber\\
{}&{}&\pp =
+
{\Omega}_-^{{\cal A}_1}(t,p)_{\alpha_1}
\dots
{\Omega}_-^{{\cal A}_n}(t,p)_{\alpha_n}
\Bigl(
f_-(t,p)^{-}_{{\cal A}_1\dots{\cal A}_n}
e^{i s\sqrt{p\cdot p}+ i p\cdot x}
+
f_-(t,p)^{+}_{{\cal A}_1\dots{\cal A}_n}
e^{-i s\sqrt{p\cdot p}- i p\cdot x}
\Bigr).
\ee
For $t=\omega$
\be
\psi_\pm(p)_{\alpha_1\dots\alpha_n}^{\dots}
&=&
(-1)^n
\left(
\begin{array}{c}
\pi_{A_1}
\dots
\pi_{A_{n-1}}
\pi_{A_n}
f_\pm(\omega,p)^{\dots}_{(-)_1\dots(-)_{n-1}(-)_n}
\\
\vdots\\
\bar \pi_{A'_1}
\dots
\pi_{A_{n-1}}
\pi_{A_n}
f_\pm(\omega,p)^{\dots}_{(+)_1\dots(-)_{n-1}(-)_n}
\\
\bar \pi_{A'_1}
\dots
\pi_{A_{n-1}}
\bar \pi_{A'_n}
f_\pm(\omega,p)^{\dots}_{(+)_1\dots(-)_{n-1}(+)_n}
\\
\bar \pi_{A'_1}
\dots
\bar \pi_{A'_{n-1}}
\pi_{A_n}
f_\pm(\omega,p)^{\dots}_{(+)_1\dots(+)_{n-1}(-)_n}
\\
\bar \pi_{A'_1}
\dots
\bar \pi_{A'_{n-1}}
\bar \pi_{A'_n}
f_\pm(\omega,p)^{\dots}_{(+)_1\dots(+)_{n-1}(+)_n}
\end{array}
\right)
+
{\rm additional \, terms}.
\label{dotted}
\ee
Each row in the ``additional terms" is composed of products of
the $\pi$'s and the $\omega$'s  and every one of the terms contains 
at least one 
$\sqrt{\frac{p\cdot p}{2}}\omega_A$ or $\sqrt{\frac{p\cdot p}{2}}\bar
\omega_{A'}$. Two conclusions follow immediately from
(\ref{dotted}). First, 
all additional terms vanish on the massless boundary.
Second, a higher-rank $\omega$-spinor is defined by
\be
\dots\omega^{A_k}\dots \bar \omega^{A'_l}\dots
\psi_\pm(p)^{\dots}_{\dots{A_k}\dots {A'_l}\dots}
&=&
f_\pm(\omega,p)^{\dots}_{\dots{(-)}\dots {(+)}\dots}
=:
f_\pm(\omega,p)^{\dots}_{\dots{\it 0}\dots {\it 1}\dots}
\label{BWn}
\ee
since the additional terms are annihilated by the transvection
with 
$\dots\omega^{A_k}\dots \bar \omega^{A'_l}\dots$. 
The part explicitly shown in (\ref{dotted}) is typical of
massless fields in both twistor \cite{PR2} and Fourier form
\cite{W}. For $p\cdot p=0$ the passive transformations discussed
in the next section are reducible and each row of (\ref{dotted})
transforms independently. 

A symmetry of a BW-spinor determines (and is determined by)
the corresponding symmetry of
$\psi_\pm(p)^{\dots}_{\dots{A_k}\dots {A'_l}\dots}$.

\subsection{On-shell amplitudes}

The on-shell amplitudes are obtained from the off-shell ones by
putting $p\cdot p=m^2$ and $p=p_m$ in the above formulas so do not
require a separate treatment. The normalization condition for
${\Omega}_\pm^{\cal A}(t,p)_{\alpha}$ used in
the definition of amplitudes does not involve any integration
and therefore is identical to the one for 
${\Omega}_\pm^{\cal A}(t,p_m)_{\alpha}$. 

\subsection{Positive-definite scalar products in momentum space in
terms of BW-spinors}

Let $f$ and $g$ denote amplitudes
corresponding to the rank-$n$ bispinors $\psi$ and 
$\phi$.
The products (\ref{off-sp}) and (\ref{on-sp}) have a simple
BW-spinor representation. 

\subsubsection{Off-shell products}

\be
\langle\psi,\phi\rangle 
&=&
\frac{1}{(2\pi)^4}\int d^4p\,
\vs{^{{\cal A}_1}}{^{{\cal B}_1}}\dots \vs{^{{\cal
A}_n}}{^{{\cal B}_n}} 
\Bigl(
\bar f_+(t,p)^+_{{{\cal A}_1\dots {\cal A}_n}}
g_+(t,p)^+_{{{\cal B}_1\dots {\cal B}_n}}
+
\bar f_+(t,p)^-_{{{\cal A}_1\dots {\cal A}_n}}
g_+(t,p)^-_{{{\cal B}_1\dots {\cal B}_n}}
\nonumber\\
&\pp =&
\pp {\frac{1}{(2\pi)^4}\int d^4p\,
\vs{^{{\cal A}_1}}{^{{\cal B}_1}}\dots \vs{^{{\cal
A}_n}}{^{{\cal B}_n}} 
\Bigl(}
+
\bar f_-(t,p)^+_{{{\cal A}_1\dots {\cal A}_n}}
g_+(t,p)^+_{{{\cal B}_1\dots {\cal B}_n}}
+
\bar f_-(t,p)^-_{{{\cal A}_1\dots {\cal A}_n}}
g_+(t,p)^-_{{{\cal B}_1\dots {\cal B}_n}}
\Bigr).
\label{BW-off-sp}
\ee

\subsubsection{On-shell products}

\be
\langle\psi,\phi\rangle_{m^2}
&=&
\frac{1}{(2\pi)^3}\int \frac{d^3p}
{2|p_m^0|}
\vs{^{{\cal A}_1}}{^{{\cal B}_1}}\dots \vs{^{{\cal
A}_n}}{^{{\cal B}_n}} 
\Bigl(
\bar f_+(t,p_m)^+_{{{\cal A}_1\dots {\cal A}_n}}
g_+(t,p_m)^+_{{{\cal B}_1\dots {\cal B}_n}}
+
\bar f_+(t,p_m)^-_{{{\cal A}_1\dots {\cal A}_n}}
g_+(t,p_m)^-_{{{\cal B}_1\dots {\cal B}_n}}
\nonumber\\
&\pp =&
\pp {\frac{1}{(2\pi)^4}\int d^4p\,
\vs{^{{\cal A}_1}}{^{{\cal B}_1}}\dots \vs{^{{\cal
A}_n}}{^{{\cal B}_n}} 
\Bigl(}
+
\bar f_-(t,p_m)^+_{{{\cal A}_1\dots {\cal A}_n}}
g_+(t,p_m)^+_{{{\cal B}_1\dots {\cal B}_n}}
+
\bar f_-(t,p_m)^-_{{{\cal A}_1\dots {\cal A}_n}}
g_+(t,p_m)^-_{{{\cal B}_1\dots {\cal B}_n}}
\Bigr).
\label{BW-on-sp}
\ee

\section{Passive transformations of $\omega$-spinors}
\label{Sec.VII}

In this section we formulate the main result of this work: The
manifestly covariant version of unitary representations in terms
of the passive transformations of $\omega$-spinors. The
general $t$-spinor form can be obtained by 
the unitary similarity transformation (\ref{calW}) and will not
be explicitly discussed. Alternatively, it could be directly
obtained with the help of (\ref{15}) and (\ref{16}). The
complicated formulas (\ref{15}) and (\ref{16}) show the scale of
simplification obtained by the choice of $t=\omega$.

\subsection{Rank-1 $\omega$-spinors}

Consider the active Poincar\'e transformations (\ref{I.91})--(\ref{I.94})
of eigenvectors of $S\bigl(\omega(p),p\bigr)$:
\be
{\Omega}_\pm^{\cal A}\bigl(\omega(p),p\bigr)_\alpha
f_\pm\bigl(\omega(p),p\bigr)^{+}_{\cal A}
&\to&
e^{\mp ip\cdot x}
S_{\alpha}{^{\beta}}{\Omega}_\pm^{\cal
A}\bigl(\omega(S^{-1}p),S^{-1}p\bigr)_\beta 
f_\pm\bigl(\omega(S^{-1}p),S^{-1}p\bigr)^{+}_{\cal A}\nonumber\\
&=&
e^{\mp ip\cdot x}
{\Omega}_\pm^{\cal
A}\bigl(S\omega(p),p\bigr)_\alpha 
f_\pm\bigl(\omega(S^{-1}p),S^{-1}p\bigr)^{+}_{\cal A}\\
&=:&
{\Omega}_\pm^{\cal
A}\bigl(\omega(p),p\bigr)_\alpha 
{\cal U}(x,S)f_\pm\bigl(\omega(p),p\bigr)^{+}_{\cal A},\\
{\Omega}_\pm^{\cal A}\bigl(\omega(p),p\bigr)_\alpha
f_\pm\bigl(\omega(p),p\bigr)^{-}_{\cal A}
&\to&
e^{\pm ip\cdot x}
S_{\alpha}{^{\beta}}{\Omega}_\pm^{\cal
A}\bigl(\omega(S^{-1}p),S^{-1}p\bigr)_\beta 
f_\pm\bigl(\omega(S^{-1}p),S^{-1}p\bigr)^{-}_{\cal A}\nonumber\\
&=&
e^{\pm ip\cdot x}
{\Omega}_\pm^{\cal A}\bigl(S\omega(p),p\bigr)_\alpha 
f_\pm\bigl(\omega(S^{-1}p),S^{-1}p\bigr)^{-}_{\cal A}\\
&=:&
{\Omega}_\pm^{\cal A}\bigl(\omega(p),p\bigr)_\alpha 
{\cal U}(x,S)f_\pm\bigl(\omega(p),p\bigr)^{-}_{\cal A}.
\ee
We have used here the definitions (\ref{38})--(\ref{I.41}), (\ref{sf3}),
(\ref{sf4}) and transformation properties (\ref{I.53}), (\ref{I.54}). The
active Poincar\'e transformations of the bispinors generate the
corresponding passive transformations of the $\omega$-spinors.
To simplify notation we shall concentrate on the nontrivial,
$SL(2,C)$ part of the Poincar\'e transformations (with $x=0$).
From now on we shall write all $\omega$-spinors
$f(\omega(p),p)$ simply as $f(p)$. This will not lead to
ambiguities since no other $t$-spinors will be considered.
The bispinor formulas can be written in terms of 2-spinors (cf.
(\ref{I.96})) as 
\be
{}&{}&
\left(
\begin{array}{c}
\pm\sqrt{\frac{p\cdot p}{2}}
\omega{_{A}}(p)\\
- \bar \pi{_{A'}}(p)
\end{array}
\right)
{\cal U}(0,S)f_\pm(p)^{\dots}_{\it 1}
+
\left(
\begin{array}{c}
- \pi{_{A}}(p)\\
\mp\sqrt{\frac{p\cdot p}{2}}
\bar \omega{_{A'}}(p)
\end{array}
\right)
{\cal U}(0,S)f_\pm(p)^{\dots}_{\it 0}\nonumber\\
&{}&=
\left(
\begin{array}{c}
\pm\sqrt{\frac{p\cdot p}{2}}
S\omega{_{A}}(p)\\
- \overline{S\pi}{_{A'}}(p)
\end{array}
\right)
f_\pm(S^{-1}p)^{\dots}_{\it 1}
+
\left(
\begin{array}{c}
- S\pi{_{A}}(p)\\
\mp\sqrt{\frac{p\cdot p}{2}}
\overline{S\omega}{_{A'}}(p)
\end{array}
\right)
f_\pm(S^{-1}p)^{\dots}_{\it 0}
\ee
and the explicit matrix form of the passive transformation is
\be
\left(
\begin{array}{c}
{\cal U}(0,S)f_\pm(p)^{\dots}_{\it 0}\\
{\cal U}(0,S)f_\pm(p)^{\dots}_{\it 1}
\end{array}
\right)
&=&
\left(
\begin{array}{cc}
\omega{_{A}}(p)S\pi{^{A}}(p) & 
\mp\sqrt{\frac{p\cdot p}{2}}\omega{_{A}}(p)
S\omega{^{A}}(p)\\
\pm\sqrt{\frac{p\cdot p}{2}}
{\bar \omega}{_{A'}}(p)\overline{S\omega}{^{A'}}(p) &
{\bar \omega}{_{A'}}(p)\overline{S\pi}{^{A'}}(p)
\end{array}
\right)
\left(
\begin{array}{c}
f_\pm(S^{-1}p)^{\dots}_{\it 0}\\
f_\pm(S^{-1}p)^{\dots}_{\it 1}
\end{array}
\right).\label{matrix}
\ee
An arbitrary passive Poincar\'e transformation of the
$\omega$-spinors can be written as 
\be
{\cal U}(x,S)f_\pm(p)^{+}_{\cal A}
&=&
e^{\mp ip\cdot x}
U_\pm(S,p){_{\cal A}}{^{\cal B}}
f_\pm(S^{-1}p)^{+}_{\cal B},
\label{full1}\\
{\cal U}(x,S)f_\pm(p)^{-}_{\cal A}
&=&
e^{\pm ip\cdot x}
U_\pm(S,p){_{\cal A}}{^{\cal B}}
f_\pm(S^{-1}p)^{-}_{\cal B}.
\label{full2}
\ee
To see that $U_\pm(S,p)\in SU(2)$ it is sufficient to denote $o_A=S\omega_A$, 
$\iota_A=S\pi_A$ and,
using  (\ref{sf1}), (\ref{sf2}) and $o_A\iota{^{B}}
-\iota_A o^B=\ve{_A}{^B}$, show that  $\det U_\pm(S,p) =
\omega_A \pi^A=1$. 
Eqs.~(\ref{full1}), (\ref{full2}) show that there are, in general,
four classes of unitary representations corresponding to the
four combinations of signs of ``energy" and ``mass". The signs
typically associated with the signs of energy (e.g. in the
on-shell version of the Dirac equation) are those occuring in
the off-diagonal elements of the transformation matrix
$U_\pm(S,p)$. The two matrices $U_\pm(S,p)$ reduce to a single,
diagonal $SU(2)$ matrix for $p\cdot p=0$ and the four representations
reduce to two. The fact that for $p\cdot p=0$ the transformation
becomes  reducible and is a direct sum of one dimensional
representations is well known in representation theory
\cite{m=0-1,m=0-2,m=0-3}. 
Massless irreducible unitary representations are typically obtained via 
induction from unitary representations of $SE(2)$ which
are either one-dimensional or infinite-dimensional. 
Our approach shows that even the massless (discrete spin) 
representations can be regarded as induced from $SU(2)$ but for the price of
reducibility which is not regarded as fundamentally important in
this work. The fact that the massless limit eliminates one of
the signs and allows for superpositions of states which are
forbidden for $m\neq 0$ resembles an analogous phenomenon of vanishing of
{\it charge\/} for particles of mass zero.

\subsection{Complex conjugated $\omega$-spinors}

The unitarity conditions (\ref{I.109}), (\ref{I.110}), and the
transformation rules (\ref{full1}), (\ref{full2})
imply that, in addition to (\ref{ccBW1}) and (\ref{ccBW2}), 
\be
\overline{f_\pm(p)_{\it 0}^+}&=& \bar f_\mp(p)_{\it 1}^-\\
\overline{f_\pm(p)_{\it 1}^+}&=& \bar f_\mp(p)_{\it 0}^-\\
\overline{f_\pm(p)_{\it 0}^-}&=& \bar f_\mp(p)_{\it 1}^+\\
\overline{f_\pm(p)_{\it 1}^-}&=& \bar f_\mp(p)_{\it 0}^+.
\ee

\subsection{Higher-rank $\omega$-spinors}

Higher-rank $\omega$-spinors are obtained by taking tensor
products of rank-1 representations. This requires no further
comments since one can apply the standard $SU(2)$-spinor methods. 

\subsection{Proof of ${\cal U}(0,S_1){\cal U}(0,S_2)={\cal U}(0,S_1S_2)$}

The passive transformations are unitary. The fact that they form
a representation follows directly from the way we obtained them:
We have started from active spinor transformations that have the
representation property. Therefore one way of
proving that ${\cal U}(x,S)$ is a representation is to
switch from the BW-spinor level to bispinors, apply the
standard 2-spinor formalism,  and then return again to
BW-spinors. It is interesting and instructive, however, to see
how the manifestly covariant spinor techniques allow to prove this
directly at the BW-spinor level without any reference to
bispinors and active spinor transformations.

We shall concentrate on the $SL(2,C)$ part of the
proof. If $\phi_A(p)$ and $\psi_A(p)$ are arbitrary spinor
fields then
\be
\phi_A(S_1^{-1}p)S_2\psi^A(S_1^{-1}p)
=
S_1\phi_A(p)S_1S_2\psi^A(p).
\ee
Applying this identity to the spin frames we obtain
\be
{}&{}&
\left(
\begin{array}{c}
{\cal U}(0,S_1)
\bigl[{\cal U}(0,S_2)f\bigr]_\pm(p)^{\dots}_{\it 0}\\
{\cal U}(0,S_1)
\bigl[{\cal U}(0,S_2)f\bigr]_\pm(p)^{\dots}_{\it 1}
\end{array}
\right)\nonumber\\
&{}& =
\left(
\begin{array}{cc}
\omega{_{A}}(p)S_1\pi{^{A}}(p) & 
\mp\sqrt{\frac{p\cdot p}{2}}\omega{_{A}}(p)
S_1\omega{^{A}}(p)\\
\pm\sqrt{\frac{p\cdot p}{2}}
{\bar \omega}{_{A'}}(p)\overline{S_1\omega}{^{A'}}(p) &
{\bar \omega}{_{A'}}(p)\overline{S_1\pi}{^{A'}}(p)
\end{array}
\right)\nonumber\\
&{}&\pp =\times
\left(
\begin{array}{cc}
S_1\omega{_{B}}(p)S_1S_2\pi{^{B}}(p) & 
\mp\sqrt{\frac{p\cdot p}{2}}S_1\omega{_{B}}(p)
S_1S_2\omega{^{B}}(p)\\
\pm\sqrt{\frac{p\cdot p}{2}}
\overline{S_1\omega}{_{B'}}(p)\overline{S_1S_2\omega}{^{B'}}(p) &
\overline{S_1\omega}{_{B'}}(p)\overline{S_1S_2\pi}{^{B'}}(p)
\end{array}
\right)
\left(
\begin{array}{c}
f_\pm((S_1S_2)^{-1}p)^{\dots}_{\it 0}\\
f_\pm((S_1S_2)^{-1}p)^{\dots}_{\it 1}
\end{array}
\right)
\ee
To complete the proof one uses the
following two sequences of identities following from
(\ref{sf1}), (\ref{sf2}):
\be
{}&{}&
\omega_AS_1\pi^A\,S_1\omega_BS_1S_2\pi^B
-
\frac{p\cdot p}{2}
\omega_AS_1\omega^A\,\overline{S_1\omega}_{B'}\overline{S_1S_2\omega}^{B'}
=
\omega_A
\bigl(S_1\pi^AS_1\omega_B-S_1\omega^AS_1\pi_B\bigr)
S_1S_2\pi^B=\omega_AS_1S_2\pi^A\\
&{}&
\omega_AS_1\pi^A\,S_1\omega_BS_1S_2\omega^B
+
\omega_AS_1\omega^A\,{S_1S_2\omega}_{B}{S_1\pi}^{B}
=
\omega_A
\bigl(S_1\pi^AS_1\omega_B-S_1\omega^AS_1\pi_B\bigr)
S_1S_2\omega^B=\omega_AS_1S_2\omega^A.
\ee
Some care is needed on the massless boundary $p\cdot p=0$
where we can
take advantage of proportionality of $\pi_A$ and $S\pi_A$
implied by (\ref{sf1}).

\subsection{Generators}

Let $J^{ab}=S^{ab}+L^{ab}$ and $P^a$ denote generators of a
spinor representation of the Poincar\'e group. Here $S^{ab}$ 
are the generators of $(1/2,0)$ or $(0,1/2)$ spinor
representations of $SL(2,C)$, $L^{ab}$ is the orbital part which
occurs independently of spin, and
$P^a$ generate space-time translations. 
The corresponding generators of the unitary representation will
be denoted by calligraphic letters:
\be
{\cal P}^a&=&\pm p^a\quad{\rm for}\,(\ref{full1}),\\
{\cal P}^a&=&\mp p^a\quad{\rm for}\,(\ref{full2}),
\ee
and
\be
{\cal J}^{ab}{_{\cal A}}{^{\cal B}}
&=&
\left(
\begin{array}{cc}
\omega{_{X}}(p)J^{ab}{^{X}}{^{Y}}\pi{_{Y}}(p) & 
\mp\sqrt{\frac{p\cdot p}{2}}\omega{_{X}}(p)
J^{ab}{^{X}}{^{Y}}\omega{_{Y}}(p)\\
\pm\sqrt{\frac{p\cdot p}{2}}
{\bar \omega}{_{X'}}(p)J^{ab}{^{X'}}{^{Y'}}\bar \omega{_{Y'}}(p) &
{\bar \omega}{_{X'}}(p)J^{ab}{^{X'}}{^{Y'}}\bar \pi{_{Y'}}(p)
\end{array}
\right)
+L^{ab}\ve{_{\cal A}}{^{\cal B}}\label{J}
\ee
for both (\ref{full1}) and (\ref{full2}). The off- and on-shell
representations differ in the forms of $L^{ab}$. In the unitary
represenation case this concerns also the BW-spinor part which
contains matrix elements of $J^{ab}$ taken between the
spin-frame spinors. The on-shell version of $L^{ab}$ depends on
the choice of the reference frame which defines $\bbox p$. 
The off-shell generators are fully manifestly covariant. Using
the explicit forms of the spinor off-shell generators
\be
J^{ab}{_{X}}{^{Y}}
&=&
i\bigl(p^a\partial^b-p^b\partial^a\bigr)\ve{_{X}}{^{Y}}
+
\frac{i}{2}\ve{^{A'}}{^{B'}}
\Bigl(
\ve{^{A}}{_{X}}\ve{^{B}}{^{Y}}
+
\ve{^{B}}{_{X}}\ve{^{A}}{^{Y}}
\Bigr),\\
J^{ab}{_{X'}}{^{Y'}}
&=&
i\bigl(p^a\partial^b-p^b\partial^a\bigr)\ve{_{X'}}{^{Y'}}
+
\frac{i}{2}\ve{^{A}}{^{B}}
\Bigl(
\ve{^{A'}}{_{X'}}\ve{^{B'}}{^{Y'}}
+
\ve{^{B'}}{_{X'}}\ve{^{A'}}{^{Y'}}
\Bigr),
\ee
and the identity (\ref{partial^b}), we obtain
\be
\omega{_{X}}J^{ab}{^{X}}{^{Y}}\pi{_{Y}}
&=&
\frac{i}{2}
\Bigl(
\ve^{A'B'}\omega^{(A}\pi^{B)}
-
\ve^{AB}\bar \omega^{(A'}\bar \pi^{B')}
\Bigr)
+
\frac{i}{2}\omega_X
\bigl(
p^a\partial^b-p^b\partial^a
\bigr)
\pi^X
-
\frac{i}{2}\bar \omega_{X'}
\bigl(
p^a\partial^b-p^b\partial^a
\bigr)
\bar \pi^{X'}
\ee
which explicitly shows that the matrix in (\ref{J}) is
Hermitian. The same argument can be used to show Hermiticity in
the on-shell version. 
The form (\ref{J}) seems to be the first manifestly covariant
version of generators of the unitary representations which can
be found in literature. A review of other explicit forms of
the generators is given in \cite{gen2}. Let me also
remark here that one of the
advantages of the $\omega$-spinor approach lies in obtaining
directly and explicitly an ``integrated" form of the
representation. It is easy to find generators once we have the
integrated representation. The opposite direction is usually much more
complicated. The typical forms of
integrated representations are given in terms of exponents 
\cite{M5,Ritus}. The $\omega$-spinor approach results in 
a form in which the (implicit) exponents are already evaluated.

\section{Off-shell fields on the Poincar\'e group}
\label{Sec.VIII}

We have briefly used the Minkowski 4-position representation to
relate the spin-energy projectors to Bargmann-Wigner wave
equations and to motivate the 4-momentum
form of active Poincar\'e transformations. The
4-position representation played a role of a formal tool
essentially void of any special physical importance.
From a group representation point of view the $x$-dependent
factors $e^{\pm i p\cdot x}$ are representations of translations
by $x$ and there is basically no reason to use only the four out of
the ten parameters of the Poincar\'e group in transition to a
``position" representation. This observation is a departure
point of the formalisms developed by Lur\c cat \cite{Lurcat} and
Toller \cite{T1,T2}. The version given below is naturally
implied by the BW-spinor formalism. Let $g=(x,S)$ be an element of the
Poincar\'e group, 
$
S=S(y)=e^{-\frac{i}{2}y^{ab}{J}_{ab}},
$
and
\be
{\cal U}\bigl(x,S(y)\bigr)=:{\cal U}(x,y)=:{\cal U}(g)
=
e^{-ix^a{\cal P}_a -\frac{i}{2}y^{ab}{\cal J}_{ab}}.
\ee
We define the {\it right\/} off-shell generalized position representation as
\be
f^R_{\dots}(x,y)^{\dots}_{\cal A}=f^R_{\dots}(g)^{\dots}_{\cal A}
=
\frac{1}{(2\pi)^4}
\int d^4p\,
{\cal U}(x,y)f_{\dots}(p)^{\dots}_{\cal A}.
\ee
The definition implies 
\be
f^R_{\dots}(g_1g_2)^{\dots}_{\cal A}={\cal
U}(g_2)f^R_{\dots}(g_1)^{\dots}_{\cal A} 
\ee
and the Poincar\'e group acts in this position space in terms of
a right regular representation. 
Analogously we define the {\it left\/} off-shell generalized position
representation as 
\be
f^L_{\dots}(x,y)^{\dots}_{\cal A}=f^L_{\dots}(g)^{\dots}_{\cal A}
=
\frac{1}{(2\pi)^4}
\int d^4p\,
{\cal U}^{-1}(x,y)f_{\dots}(p)^{\dots}_{\cal A}.
\ee
which implies
\be
f^L_{\dots}(g^{-1}_1g_2)^{\dots}_{\cal A}={\cal
U}(g_1)f^L_{\dots}(g_2)^{\dots}_{\cal A} 
\ee
and the group acts in terms of its left regular representation. 

The off-shell scalar products have the following position
representation 
\be
{}&{}&
\int d^4x\,
\vs{^{{\cal A}_1}}{^{{\cal B}_1}}\dots \vs{^{{\cal
A}_n}}{^{{\cal B}_n}} 
\Bigl(
\bar f_+(x,y)^+_{{{\cal A}_1\dots {\cal A}_n}}
g_+(x,y)^+_{{{\cal B}_1\dots {\cal B}_n}}
+
\bar f_+(x,y)^-_{{{\cal A}_1\dots {\cal A}_n}}
g_+(x,y)^-_{{{\cal B}_1\dots {\cal B}_n}}
\nonumber\\
&{}&
\pp {\frac{1}{(2\pi)^4}\int d^4p\,
\vs{^{{\cal A}_1}}{^{{\cal B}_1}}\dots \vs{^{{\cal
A}_n}}{^{{\cal B}_n}} 
\Bigl(}
+
\bar f_-(x,y)^+_{{{\cal A}_1\dots {\cal A}_n}}
g_+(x,y)^+_{{{\cal B}_1\dots {\cal B}_n}}
+
\bar f_-(x,y)^-_{{{\cal A}_1\dots {\cal A}_n}}
g_+(x,y)^-_{{{\cal B}_1\dots {\cal B}_n}}
\Bigr)\nonumber\\
{}&{}&=
\frac{1}{(2\pi)^4}\int d^4p\,
\vs{^{{\cal A}_1}}{^{{\cal B}_1}}\dots \vs{^{{\cal
A}_n}}{^{{\cal B}_n}} 
\Bigl(
\bar f_+(p)^+_{{{\cal A}_1\dots {\cal A}_n}}
g_+(p)^+_{{{\cal B}_1\dots {\cal B}_n}}
+
\bar f_+(p)^-_{{{\cal A}_1\dots {\cal A}_n}}
g_+(p)^-_{{{\cal B}_1\dots {\cal B}_n}}
\label{spx}\\
&\pp =&
\pp {\frac{1}{(2\pi)^4}\int d^4p\,
\vs{^{{\cal A}_1}}{^{{\cal B}_1}}\dots \vs{^{{\cal
A}_n}}{^{{\cal B}_n}} 
\Bigl(}
+
\bar f_-(p)^+_{{{\cal A}_1\dots {\cal A}_n}}
g_+(p)^+_{{{\cal B}_1\dots {\cal B}_n}}
+
\bar f_-(p)^-_{{{\cal A}_1\dots {\cal A}_n}}
g_+(p)^-_{{{\cal B}_1\dots {\cal B}_n}}
\Bigr)
\ee
and we automatically avoid the well known 
problems with the position representation of the ordinary
on-shell fields. Notice that we do not integrate over $y$ as the
4-position probability density is $y$-independent so that the
$y$'s play a role of internal degrees of freedom. The pair
$(x,y)$ plays a role of an extended configuration space consisting of
time, position, velocity and angles. This should not be confused
with various phase-spaces associated with frames and wavelets
\cite{Kjmp1,Kjmp2,Kjmp3,K1,K2,Prugovecki}. 
Scalar product (\ref{spx}) is positive-definite and is a natural
alternative to the indefinite metric used in the
off-shell approaches to the Dirac equation
\cite{arg1,arg2,arg3,Gos}. This kind of product was
proposed by Horwitz and Piron in \cite{H1} but their wave functions
were zero-spin and $y$-independent. 
Local expressions can be obtained also in the on-shell theory
but one has to consider analytically continued
fields and wavelets~\cite{K1,K2}.

\section{Summary and comments}

We have started with the field of spin-frames ($\omega_A(p),\,\pi_A(p)$)
associated with $p$, and used them to
simplify the eigenvalue problem for the P-L vector projection in
a direction given by a world-vector $t$. As opposed to the
standard treatments where $t$ is a time direction (the same for
all $p$'s) our $t$ is, in general, $p$-dependent and timelike,
spacelike or null. The
corresponding eigenstates play a role of a basis used to
define the (Bargmann-Wigner) amplitudes. 
The BW-amplitudes are what one usually calls the {\it noncovariant
Wigner states\/} obtained via induction from little groups of fixed
4-momenta. Our construction does not use the induction procedure,
is manifestly covariant and in addition works simultaneously
for both massive and massless cases. 
The case of imaginary mass can be formulated in
an analogous way but for technical reasons it has to be treated
separately so we do not do it here. 
The amplitudes transform as
scalars under active $SL(2,C)$ transformations. From the
viewpoint of passive transformations the amplitudes are local $SU(2)$ spinors
and for this reason we term them the BW-spinors. 
Of particular interest are special BW-spinors ($\omega$-spinors)
which are associated with the flagpole directions of the
spin-frame field $\omega_A(p)$. We show by an explicit
spinor calculation that the unitary passive transformations form
a represenation. We explicitly find its generators and discuss a
generalized off-shell position representation in terms of fields
on the Poincar\'e group. 
Although we occasionally use the numerical BW-indices to make
objects such as $\vs$ explicit, the whole
construction can be understood also as an   
abstract index one. 
To be able to do this we first had to introduce a bispinor index notation
which is somehow in-between the familiar 2-spinor and twistor
conventions.

The fact that our $\omega$-spinors resemble in many respects the
spin-weighted spherical harmonics of Newman and Penrose
\cite{PR} can be used to find the $\omega$-spinor version of
harmonics but this will not be discussed in this paper. 
The problem of surface harmonics formulation of the unitary
representations is described in the works of Moses \cite{M4} 
but his version does not satisfy our standards of manifest
covariance.

\section{Acknowledgements}

Parts of this work were done during my stays at Massachusetts
Institute of Technology, University
of Massachusetts at Lowell and Oaxtepec, Mexico. I'm indebted to
David~E. Pritchard, Gerald Kaiser and Bogdan Mielnik for
hospitality, support and exeptional scientific atmosphere they
managed to create.  
I gratefully acknowledge many hours of extensive and
inspiring discussions on the subject with G.~Kaiser.

\section{Appendices}

\subsection{Bispinor abstract index convention}
\label{bispinors}

The convention I use is the following. To any Greek
abstract index there corresponds a pair of Latin ones written
down in a lexicographic order. For example
\be
F{_{\alpha}}{^{\beta'}}{_{\gamma}}=
\left(
\begin{array}{c}
F{_{A}}{^{B'}}{_{C}}\\
F{_{A}}{^{B'}}{_{C'}}\\
F{_{A}}{^{B}}{_{C}}\\
F{_{A}}{^{B}}{_{C'}}\\
F{_{A'}}{^{B'}}{_{C}}\\
\vdots
\end{array}
\right);
\quad
{\ve}_{\alpha}{^{\beta'}}=
\left(
\begin{array}{c}
{\ve}_{A}{^{B'}}\\
{\ve}_{A}{^{B}}\\
{\ve}_{A'}{^{B'}}\\
{\ve}_{A'}{^{B}}
\end{array}
\right)=
\left(
\begin{array}{c}
0\\
\ve_{A}{^{B}}\\
\ve_{A'}{^{B'}}\\
0
\end{array}
\right).\nonumber
\ee
Bispinors of any rank are written as columns. This concerns also
Dirac gamma matrices and spin and energy projectors which
normally would be written in a matrix form.
Any permutation preserving the lexicographic rule induces a
natural isomorphism, say, $F{_{\alpha}}{^{\beta'}}{_{\gamma}}
\to F{_{\alpha'}}{^{\beta'}}{_{\gamma}}$ where the latter 
bispinor would begin with $F{_{A'}}{^{B'}}{_{C}}$. 
In particular 
\be
\psi_\alpha=
\left(
\begin{array}{c}
\psi_A\\
\psi_{A'}
\end{array}
\right),
\quad
\bar \psi_{\alpha'}=
\left(
\begin{array}{c}
\bar \psi_{A'}\\
\bar \psi_{A}
\end{array}
\right),
\quad
\bar \psi_{\alpha}=
\left(
\begin{array}{c}
\bar \psi_{A}\\
\bar \psi_{A'}
\end{array}
\right).
\ee
The bispinor summation convention is illustrated by
$G^\alpha H_\alpha=G^{A} H_{A}+G^{A'} H_{A'}=G^{\alpha'} H_{\alpha'}$.

\subsection{General properties of spin-frames associated with 4-momenta}

Consider two spin-frame fields $\bigl(\omega_A(p),\pi_A(p)\bigr)$ and
$\bigl(o_A(p),\iota_A(p)\bigr)$ satisfying (\ref{p})
\be
p^a=
\pi^{A}(p) \bar \pi^{A'}(p)
+
\frac{p\cdot p}{2}
\omega^{A}(p) \bar \omega^{A'}(p)
=
\iota^{A}(p) \bar \iota^{A'}(p)
+
\frac{p\cdot p}{2}
o^{A}(p) \bar o^{A'}(p)
.\label{p2}
\ee
Transvecting (\ref{p2}) with $\pi_A(p)\bar o_{A'}(p)$ or $\omega_A(p)\bar
o_{A'}(p)$ we obtain 
\be
\pi_{A}(p)\iota^A(p)&=&\frac{p\cdot p}{2}
\bar \omega_{A'}(p)\bar o^{A'}(p),\label{sf1}\\
\omega_A(p)\iota^A(p)&=&\bar o_{A'}(p)\bar \pi^{A'}(p).\label{sf2}
\ee
Eq.~(\ref{p2}) implies that the field of spin-frames is a spinor field,
i.e. the pair $\bigl(S\omega_A(p),S\pi_A(p)\bigr)$
where 
\be
S\omega_A(p)&=&S{_A}{^B}\omega_B(S^{-1}p)\label{sf3}\\
S\pi_A(p)&=&S{_A}{^B}\pi_B(S^{-1}p)\label{sf4}
\ee
is a spin-frame satisfying (\ref{p2}). 

Transvecting (\ref{p2}) with $\pi_A(p)$ or $\omega_A(p)$ we find that
\be
\pi_A(p)p^{AA'} &=& -\frac{p\cdot p}{2}\bar \omega^{A'}(p),\\
\omega_A(p)p^{AA'} &=& \bar \pi^{A'}(p),
\ee
which imply ($\partial^b=\partial/\partial p_b$)
\be
\omega_A(p)\partial^b\pi^A(p) + \bar \omega_{A'}(p)\partial^b\bar
\pi^{A'}(p)= 
\omega^B(p)\bar \omega^{B'}(p).\label{partial^b}
\ee

\subsection{Explicit example of decomposition of $p$ in terms of
spin-frames}\label{spin-frame}

Consider an arbitrary $p$-independent
spinor $\nu^A\neq 0$. 
Let $\omega^{a}=\omega^{A}\bar \omega^{A'}$, 
$\pi^a=\pi^A \bar \pi^{A'}$, where
\be
\omega^{A}
&=&
\frac{\nu^A}{
\sqrt{p^{BB'}\nu_B
\bar \nu_{B'}}}=\omega^{A}(\nu,p)\label{oo}\\
\pi^A 
&=&
\frac{p^{AA'}\bar \nu_{A'}}
{\sqrt{p^{BB'}\nu_B
\bar \nu_{B'}}}=\pi^A(\nu, p).\label{pp}
\ee
These spin frames are defened globally for timelike $p$ since
$p\cdot \nu$ never vanishes. For null $p$ we can use 
\be
\pi^A(\nu, p)
&=&
\frac{p^{AA'}\bar \nu_{A'}}
{\sqrt{p^{BB'}\nu_B
\bar \nu_{B'}}},\label{ppp}\\
\omega^{A}(\nu,n,p)
&=&-
\frac{n^{AA'}\bar \pi_{A'}(\nu,p)}
{n^a\,p_a},
\label{ooo}
\ee
where $n$ is timelike. The spin-frames satisfy (\ref{p}).

\end{document}